\documentclass[smallcondensed]{svjour3}

\usepackage[a4paper,margin=3cm]{geometry}

\usepackage{hyperref,graphicx,color,fancyhdr}
\usepackage[caption=false,subrefformat=parens,labelformat=empty]{subfig}

\usepackage{amsmath,amssymb,mathrsfs,amsfonts}
\usepackage{cases}
 
\journalname{Journal of Statistical Physics}

\newcommand{\der}[2]{\frac{d #1}{d #2}}
\newcommand{\pd}[2]{\frac{\partial #1}{\partial #2}}

\let\temp\epsilon
\let\epsilon\varepsilon
\let\varepsilon\temp

\begin{document}

\title{Localization in the Discrete Non-Linear Schr\"odinger Equation and geometric properties of the microcanonical surface}
\titlerunning{Localization in the DNLSE and geometric properties of the microcanonical surface}        % if too long for running head

\author{Claudio Arezzo$^{1,2}$ \and Federico Balducci$^{1,3,4,*}$ \and Riccardo Piergallini$^{5}$ \and Antonello Scardicchio$^{1,3}$ \and Carlo Vanoni$^{3,4}$ }
\authorrunning{C. Arezzo, F. Balducci, R. Piergallini, A. Scardicchio, and C. Vanoni} % if too long for running head

\institute{$^1$ The Abdus Salam ICTP, Strada Costiera 11, 34151, Trieste, Italy  \\
$^2$ Universit\`a di Parma -- Dipartimento di Scienze Matematiche, Fisiche e Informatiche, Parco Area delle Scienze, 53/A 43124 Parma, Italy \\
$^3$ INFN Sezione di Trieste, Via Valerio 2, 34127 Trieste, Italy\\
$^4$ SISSA, via Bonomea 265, 34136, Trieste, Italy\\
$^5$ Universit\`a di Camerino -- Scuola di Scienze e Tecnologie, Via Madonna delle Carceri, 62032 Camerino, Italy\\
$^*$ \email{fbalducc@sissa.it} }

%\institute{C. Arezzo \at Universit\`a di Parma -- Dipartimento di Scienze Matematiche, Fisiche e Informatiche, Parco Area delle Scienze, 53/A 43124 Parma, Italy
%\and
%C. Arezzo \and F. Balducci \and A. Scardicchio \at The Abdus Salam ICTP, Strada Costiera 11, 34151, Trieste, Italy 
%\and
%F. Balducci \and A. Scardicchio \at INFN Sezione di Trieste, Via Valerio 2, 34127 Trieste, Italy
%\and 
%R. Piergallini \at Universit\`a di Camerino -- Scuola di Scienze e Tecnologie, Via Madonna delle Carceri, 62032 Camerino, Italy
%\and 
%F. Balducci \and C. Vanoni \at SISSA, via Bonomea 265, 34136, Trieste, Italy}

\date{\today}

%------------------------------------------ABSTRACT------------------------------------------%  
\maketitle

\begin{abstract}
    It is well known that, if the initial conditions have sufficiently high energy density, the dynamics of the classical Discrete Non-Linear Schr\"odinger Equation (DNLSE) on a lattice shows a form of breaking of ergodicity, with a finite fraction of the total charge accumulating on a few sites and residing there for times that diverge quickly in the thermodynamic limit. In this paper we show that this kind of localization can be attributed to some geometric properties of the microcanonical potential energy surface, and that it can be associated to a phase transition in the lowest eigenvalue of the Laplacian on said surface. We also show that the approximation of considering the phase space motion on the potential energy surface only, with effective decoupling of the potential and kinetic partition functions, is justified in the large connectivity limit, or fully connected model. In this model we further observe a synchronization transition, with a synchronized phase at low temperatures. 
\end{abstract}

%------------------------------------------INTRO------------------------------------------%  

\section{Introduction}

In recent years the interest in non-ergodic states of matter has grown considerably, in particular, but not only, following recent theoretical and experimental developments in the study and control of many-body quantum systems. The discovery of Many-Body Localization (MBL) \cite{basko2006metal,gornyi2005interacting,oganesyan2007localization,vznidarivc2008many,de2013ergodicity,luitz2015many,nandkishore2015many} has extended the phenomenon of Anderson Localization \cite{anderson1958absence} to interacting systems and suggested the emergence of a dynamical phase characterized by local integrals of motion \cite{serbyn2013local,huse2014phenomenology,ros2015integrals,imbrie2016diagonalization,Imbrie2016Many,imbrie2017local} in disordered quantum systems. These studies have shown a potential big impact on quantum technologies, in the realm of mesoscopic quantum systems. Further extensions of the original idea have put forward the possibility that MBL-like physics could be observed in Josephson junctions chains \cite{pino2016nonergodic,pino2017multifractal}, when the initial state presents sufficiently large charge fluctuations, playing the role of quenched disorder in an otherwise clean system. Slow dynamics in such clean quantum systems (see also the works on \emph{quantum scars} \cite{turner2018weak}), appears in a form similar to the \emph{weak ergodicity breaking} characterizing spin glasses \cite{bouchaud1992weak,cugliandolo1993analytical,cugliandolo1994out,franz1994off,Kurchan1996Phase}, configurational glasses \cite{Angell2000Relaxation,Cavagna2009Supercooled,Berthier2011Theoretical}, and, in particular, non-linear oscillators models as the \emph{Fermi-Pasta-Ulam-Tsingou model} \cite{fermi1955,cretegny1998localization,berman2005fermi} and the \emph{Discrete Non-Linear Schr\"odinger Equation} (DNLSE) \cite{Kevrekidis2009}. The latter describes, among other things, the physics of Bose-Einstein condensates in optical lattices (in the semiclassical regime). In some experiments \cite{eiermann2004bright,bloch2008many} it has been observed that, in a one-dimensional lattice, Rubidium atoms survive very close to the initially prepared, localized configurations. Additionally, in numerical simulations one observes very long-lived breather-like excitations \cite{Rumpf2004Simple,Rumpf2007Growth,Rumpf2008Transition,Rumpf2009Stable,Iubini2013Discrete,Iubini2014Coarsening,Eckmann2018Breathers}, self-localization \cite{Hennig2013Nature,DeRoeck2015Asymptotic,Kruse2017Self} and, in general, weak ergodicity breaking \cite{Flach2018Weakly,Iubini2019Dynamical,gotti2020finitesize}. 

An explanation for these behaviors has been proposed, which is based on the inequivalence of the microcanonical and canonical Gibbs ensemble at high energy density \cite{Rasmussen2000Statistical,Gradenigo2021Localization,Gradenigo2021Condensation,Cherny2019NonGibbs}. We show, building on those papers, that in the limit of large connectivity even a diffusive dynamics on the microcanonical surface takes a time exponentially long in the system size, for the system to equilibrate. The limit of large connectivity, or \emph{mean-field} limit, is known to be a very good approximation in many statistical mechanics problems, and it is (in the description of phases and transition between them) \emph{exact} above a certain critical dimension. We therefore believe that our results qualitatively describe the experimentally relevant situation of up to three-dimensional lattices.

Another reason for our interest were the similarities between the quantum MBL phenomenon and the \emph{classical} ergodicity breaking phenomena in the DNLSE. Hence, we have set to investigate the origins of the latter, in particular to highlight similarities and differences. We stress again that we will focus on \emph{clean} systems with $\hbar \equiv 0$. MBL cannot survive in the semiclassical limit, as Anderson localization itself cannot; so the origin of the phenomena are definitely not the same. For the DNLSE, the interplay between disorder, quantum-mechanical localization and nonlinear effects has been the subject of vast research (see e.g.\ \cite{basko2011weak,DeRoeck2019Glassy,Kati2020Density} and references therein), but we stress that the physics behind it is appreciably different from the one discussed here.

As recognized in previous works, and in particular in \cite{Rasmussen2000Statistical,Gradenigo2021Localization,Gradenigo2021Condensation,Cherny2019NonGibbs}, the localization phenomenon at high energy density is due to entropic effects. As the main results of this study, we find \emph{first}, that the equilibration time at high energy density is exponential in $N$, \emph{second} that this is not due to a breakdown of connectivity of the topology of the microcanonic surface, where only the potential energy is taken into account, but \emph{third}, that the origin of the localization phenomenon at high energy density can be traced to the behaviour of the gap of the Laplace operator on the $(N-2)$-dimensional microcanonical energy manifold. In order to prove this we proceed with the following steps.

The first issue is about the possibility to neglect the term in the Hamiltonian containing phase variables (which allows the exchange of charges between different sites) when studying the microcanonical surface, a customary step in the literature. This is a fundamental point, since the partition function does not factorize in a form $Z=Z_{\mathit{momentum}}Z_{\mathit{position}}$, as it happens for example when looking at gases or liquids, where typically the phase-space variables $(p,x)$ appear each in its own term: $H(p,x) = K(p) + V(x)$. We therefore show that expansion around infinite temperature of the free energy (and, consequently, of all relevant observables) gets contributions from the kinetic term only at $O(1/\kappa)$, where $\kappa$ is the connectivity of the graph on which the DNLSE is considered (see Eqs.~\eqref{eq:Hamiltonian}--\eqref{eq:Hamiltonian_qphi} to fix the notation). Therefore, if one considers a fully connected model, the assumption of neglecting the kinetic term is completely justified. As said before, in the spirit of mean-field theory, this is a first approximation to the physics of finite connectivity lattices.

We solve the fully connected model finding that the ``infinite temperature phase", in which the free energy becomes essentially given by the potential term alone, extends all the way down to a finite temperature $T_s=2g$ ($g$ is the strength of the kinetic term and $v$ that of the potential). At temperatures lower than $T_s$, or equivalently energy densities $\epsilon < \epsilon_s = 1.481\dots$ (with the parameters $g=1$, $v=2$ used throughout this paper), the model enters a \emph{synchronized phase} in which the phases $\phi_i$ (see Eqs.~\eqref{eq:psi_to_qphi} and \eqref{eq:Hamiltonian_qphi}) stop rotating independently from each other and eventually move together at $T=0$ (energy density $\epsilon_{\mathit{GS}} = v/2 - 2 g = -1$). Conversely, for temperatures higher than $T_s$ the motion of the phases $\phi_i$ is incoherent and the charges move randomly on the microcanonical, potential energy surface.

Subsequently, after having highlighted the importance of the potential energy surface, we study the topology of such manifold. We prove, using both stratified Morse theory and a more direct geometrical approach, that the manifold undergoes a series of critical points (critical in the language of Morse theory, not of statistical physics) but it remains connected until energy densities $\epsilon= vN/4 =N/2$, which means \emph{super-extensive} energy. The infinite-temperature localization phase transition (previously studied in \cite{Rasmussen2000Statistical,Johansson2004Statistical,Rumpf2004Simple,Rumpf2007Growth,Rumpf2008Transition,Rumpf2009Stable,Samuelsen2013Statistical,Iubini2013Discrete,Iubini2014Coarsening,DeRoeck2015Asymptotic,Iubini2019Dynamical}), taking place at energy density $\epsilon_c = v = 2$, therefore, is \emph{not} due to a breakdown of connectivity in such manifold. Rather, we attribute it to the change in the scaling with $N$ of the smallest, non-zero eigenvalue $\lambda_1$ of (minus) the Laplace operator on the microcanonical surface (the smallest eigenvalue $\lambda_0=0$ corresponds to the uniform distribution on the manifold). Namely, for $\epsilon<\epsilon_c=2$ (the numerics agrees with the thermodynamic calculation within errors) we have $\lambda_1=O(1)$, while for $\epsilon>\epsilon_c$ we have $\lambda_1\sim e^{-\gamma N}$. The function $\gamma(\epsilon)\geq 0$ and vanishes as $\epsilon \to \epsilon_c^+$ with critical exponent close to $2$. 

We conjecture that this transition is related to an entropic effect for the motion of a particle on the microcanonical energy surface. In other words, the volume of the regions of phase space close to an imbalanced configuration (i.e.\ when a few charges are considerably larger than the others) becomes bigger, at energies $\epsilon > \epsilon_c$, than the volume of balanced configurations. We suggest that this mechanism, and the link with the behaviour of the smallest eigenvalue of the Laplacian on the microcanonical surface, is generic for DNLSE with different choices of graphs and potentials (as also indicated by the results of \cite{Johansson2004Statistical,Samuelsen2013Statistical}). \medskip

\begin{figure}
    \centering
    \includegraphics[width=0.55\columnwidth]{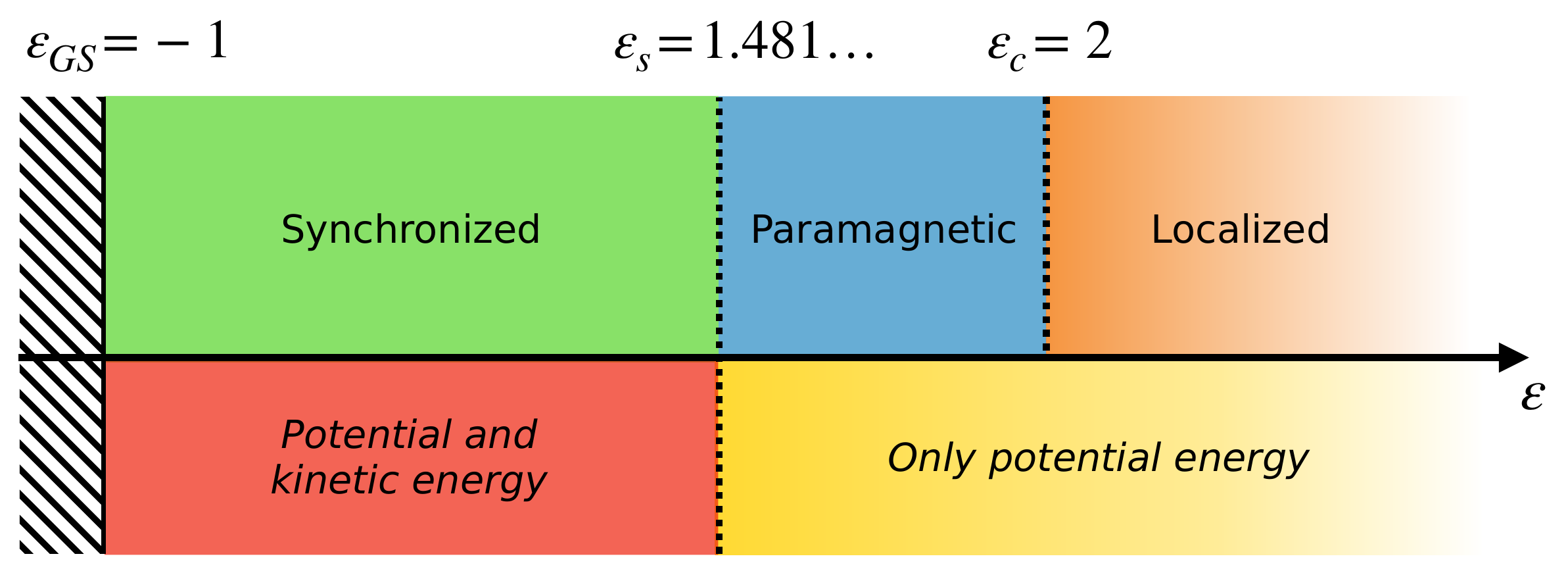}
    \caption{Phase diagram of the fully connected DNLS model for $v=2$ and $g=1$. The region $\epsilon>\epsilon_c$ corresponds to non-positive temperatures and localized dynamics; the region $\epsilon_s < \epsilon < \epsilon_c$ to ergodic incoherent dynamics for the phases $\phi_i$ (Eq.~\eqref{eq:psi_to_qphi}); the region $\epsilon < \epsilon_s$ to coherent dynamics for the same phases.}
    \label{fig:phase_diagram}
\end{figure}

The paper is organized as follows.
In Sec.\ \ref{sec:model} we introduce the DNLS model and briefly review some known results.
In Sec.\ \ref{sec:infinite_temperature} we set up an high-temperature expansion and show that, for any dimensionality, the infinite temperature point corresponds always to $\epsilon_c = v = 2$.
In Sec.\ \ref{sec:large_connectivity} we perform instead an expansion in the kinetic term of the Hamiltonian, and prove that, for large connectivity, hopping can be completely neglected in a finite neighbourhood of $\epsilon = \epsilon_c$.
In Sec.\ \ref{sec:fully_connected} we inspect more closely the reasons why hopping is sub-leading, finding out that there is a synchronization phase transition in the fully connected model at $\epsilon_s = 1.481\dots$. We discuss in detail the implications of such phase transition. 
In Sec.\ \ref{sec:topology} we describe (almost) rigorously the topology of the potential energy surface, leaving to App.\ \ref{app:topology} the flawless, yet less insightful proof. Also, we provide some intuitive explanation of the connection of the geometry with the behavior of the gap of the Laplacian.
In Sec.\ \ref{sec:dynamics} we switch to the numerical study of the dynamics of the model: we provide strong evidence that at $\epsilon = 2$ also a dynamical phase transition takes place, thus implying that the gap of the Laplacian on the potential energy surface closes as described above.
Finally, in Sec.\ \ref{sec:conclusions} we discuss the implications of our findings and speculate on future directions.

%------------------------------------------MODEL------------------------------------------%  

\section{The model}
\label{sec:model}

In this work, we want to show that the mechanism of ergodicity breaking at high energy density for the DNLSE is very general, and depends only on the particular form of the potential energy and charge conservation laws. For this reason, we consider the DNLS model on an arbitrary, regular graph $\mathcal{G}$:
\begin{equation}
    \label{eq:Hamiltonian}
    H = - \frac{g}{\kappa} \sum_{i,j=1}^N A_{ij} \left( \psi^*_i \psi^{\phantom{*}}_j + \psi^*_j \psi^{\phantom{*}}_i \right) + \frac{v}{2} \sum_{i=1}^N |\psi_i|^4.
\end{equation}
Here, the $\psi_i,\psi_i^*$ are complex fields that live on the vertices $i$ of $\mathcal{G}$, and are canonically conjugated: their Poisson brackets read $\{ \psi_i^*, \psi_j \} =i \delta_{ij}$. Then, $g$,$v$ are non-negative parameters, which we will eventually set to $g=1$ and $v=2$; for the time being, however, it is convenient to allow them to vary. Finally, $\kappa$ is the connectivity of $\mathcal{G}$ and $A$ its adjacency matrix, so each entry $A_{ij}$ is either 0 ($ij$ disconnected) or 1 ($ij$ connected). Notice that, thanks to the $1/\kappa$ normalization of the kinetic term, $H = O(N)$ for any $\mathcal{G}$, even in the limit of fully connected graph $\kappa\sim N\to\infty$, without having to rescale $g$.

Apart from the energy, there is another natural conservation law to take into account: defining the \emph{charge} $Q := \sum_i |\psi_i|^2$, it holds $\{Q, H \} = 0$. Without loss of generality, we choose to work with $Q \equiv N$ fixed from now on (or, more generally, with the average charge fixed).

We define the energy density $\epsilon := H /N$. Previous works \cite{Rasmussen2000Statistical,Gradenigo2021Localization,Gradenigo2021Condensation} have shown that $\epsilon = \epsilon_c=v$ corresponds to the $T=\infty$ limit of the model on a $1d$ chain when coupled to a thermal reservoir at temperature $T$, and that at $\epsilon_c$ the Gibbs distribution ceases to be valid: the states with energy density $\epsilon>\epsilon_c$ remain well-defined only in the microcanonical ensemble. Moreover, other works have shown that the dynamics of a chain ceases to be ergodic in this non-Gibbs phase, with the charges localizing on isolated sites in solitons rather than moving around. This has been seen both by using a simplified stochastic evolution algorithm, for all $\epsilon\geq \epsilon_c$ \cite{Iubini2013Discrete,Iubini2014Coarsening}, and with Hamiltonian dynamics \cite{Flach2018Weakly} (although in this latter work the threshold is put at $\epsilon \simeq 1.25 \, v$).

It is important to note, however, that in the proof of the canonical/microcanonical inequivalence \cite{Gradenigo2021Localization,Gradenigo2021Condensation} the kinetic term plays no role so, in particular, no role is played by the geometry of the lattice (or graph) on which the DNLSE is set. One wonders then whether the same happens at any finite temperature and finite dimensional lattice, that is to say in a left neighbourhood of $\epsilon=\epsilon_c$ and $\kappa\to\infty$. We will prove that it does, at least working at order $1/\kappa$.

%------------------------------------------INFINITE-TEMPERATURE------------------------------------------%  

\section{Infinite temperature limit}
\label{sec:infinite_temperature}

We now start exploring the limit $T \to \infty$. We start with computing the canonical partition function of the Hamiltonian \eqref{eq:Hamiltonian} which, assuming ergodicity, should tell us about the behavior of the system for $\epsilon\leq 2$. For later convenience, we employ a dimensionless chemical potential $\mu$:
\begin{equation*}
    Z(N, \beta, \mu) = \int [d\psi \, d\psi^*] e^{-\beta H + \mu Q },
\end{equation*}
where $[d\psi \, d\psi^*] := \prod_i \frac{1}{2\pi} d\psi_i \, d\psi^*_i$ and $\beta := 1/T$ (fixing $k_B \equiv 1$). First of all, we perform the canonical change of variables
\begin{equation}
    \label{eq:psi_to_qphi}
    \begin{cases}
        \psi_i = \sqrt{q_i}\, e^{i\phi_i} \\
        \psi_i^* = \sqrt{q_i} \,e^{-i\phi_i}
    \end{cases}
\end{equation}
with $q_i \geq 0$ and $\phi_i \in [0,2\pi]$. The measure becomes $[d\psi \, d\psi^*]= \prod_{i=1}^N \frac{1}{2\pi} dq_i \, d\phi_i =: [dq \, d\phi]$, the charge $Q = \sum_{i=1}^N q_i$ and the Hamiltonian 
\begin{equation}
    \label{eq:Hamiltonian_qphi}
    H = -\frac{2g}{\kappa} \sum_{i,j=1}^N A_{ij} \sqrt{q_i q_j} \cos(\phi_i - \phi_j) + \frac{v}{2} \sum_{i=1}^N q_i^2.
\end{equation}
Let us denote the thermal average of an observable $A$ as
\begin{equation*}
    \langle A \rangle_{\beta,\mu} := \frac{1}{Z} \int [dq \, d\phi]\, e^{-\beta H+\mu Q} A .
\end{equation*}
Then one can inspect the infinite temperature limit, $\beta\to 0$, by considering the expansion
\begin{equation}
    \langle A \rangle_{\beta,\mu} = \langle A \rangle_{0,\mu} - \beta \Big[\langle A H \rangle_{0,\mu} 
    - \langle A \rangle_{0,\mu} \langle H \rangle_{0,\mu} \Big] + O(\beta^2).
\end{equation}
The first thing to do is to adjust the chemical potential to have a fixed average charge (recall our choice in Sec.\ \ref{sec:model}):
\begin{align*}
    N \equiv \langle Q\rangle_{\beta,\mu} 
    &\simeq \langle Q \rangle_{0,\mu} - \beta \Big[\langle Q H \rangle_{0,\mu}- \langle Q \rangle_{0,\mu} \langle H \rangle_{0,\mu} \Big] \\
    &\simeq N \langle q \rangle_{0,\mu} - \beta \Big[ \frac{v}{2} \big(N \langle q^3 \rangle_{0,\mu} + N (N-1) \langle q \rangle_{0,\mu} \langle q^2 \rangle_{0,\mu} \big) - N \langle q \rangle_{0,\mu} \frac{v}{2} N \langle q^2 \rangle_{0,\mu} \Big] \\
    &\simeq -\frac{N}{\mu} + \beta \frac{2 N v}{\mu^3}
\end{align*}
so that $\mu \simeq -1 + 2 \beta v$. In the computation we have used the fact that the averages involving the kinetic energy vanish by symmetry, and $\langle q^k \rangle_{0,\mu} = k! / (-\mu)^k$. 

Now we focus on the internal energy. We already have found $\langle H \rangle_{0,\mu} = N v/\mu^2$; thus we just need $\langle H^2 \rangle_{0,\mu}$. The only non-zero angular integrals that figure in $\langle H^2 \rangle_{0,\mu}$ are
\begin{equation*}
    \int [d\phi]\, \cos(\phi_i - \phi_j) \cos(\phi_k - \phi_l) = \frac{1}{2} (\delta_{ik} \delta_{jl} + \delta_{il} \delta_{jk});
\end{equation*}
therefore we find
\begin{equation*}
    \langle H^2 \rangle_{0,\mu} = \frac{4 g^2}{\kappa^2} N \kappa \langle q \rangle_{0,\mu}^2 + \frac{v^2}{4} \big[ N \langle q^4 \rangle_{0,\mu} + N (N-1) \langle q^2 \rangle_{0,\mu}^2 \big]
\end{equation*}
where we recall $\kappa$ is the connectivity of the graph. The final result is
\begin{equation}    
    \label{eq:epsilon_1st_order}
    \epsilon(\beta) = v - \beta \left( v^2 + \frac{4 g^2}{\kappa} \right) + O(\beta^2)
\end{equation}
From this expression we see that the result of the non-interacting case $\epsilon(\beta=0) = v$ is not modified by the presence of the hopping on any graph geometry. We notice also that the kinetic energy term (measured by $g$) contributes only with $O(g^2/\kappa)$ and therefore vanishes to this order in the mean-field, fully connected limit $\kappa\to\infty$. We explore such limit in the next section.

%------------------------------------------LARGE-CONNECTIVITY------------------------------------------%  
\section{Large connectivity limit}
\label{sec:large_connectivity}

As noted at the end of the last section, in Eq.~\eqref{eq:epsilon_1st_order} the $O(\beta)$ correction to the internal energy density becomes independent of $g$ in the limit of large connectivity $\kappa \to \infty$. Indeed, one can verify that \emph{all the terms} in the expansion involving the hopping are subleading in $\kappa$. The situation is reminiscent of the Thouless-Anderson-Palmer (TAP) high temperature expansion of the Sherrington-Kirkpatrick model \cite{TAP}. Alongside with TAP, we can expand the free energy density $f:= - \log(Z)/ \beta N$ in powers of $1/\kappa$:
\begin{equation}
    \label{eq:TAP_expansion}
    f = f_0 + \frac{1}{\kappa} f_1 + O\left( \frac{1}{\kappa^2} \right),
\end{equation}
where $f_0$ is the free energy density at $\kappa\to\infty$, or $g=0$, while we can express $f_1$ (and successive orders too) as a sum of diagrams. 

\begin{figure}[t]
    \centering
    \includegraphics[width=0.9\textwidth]{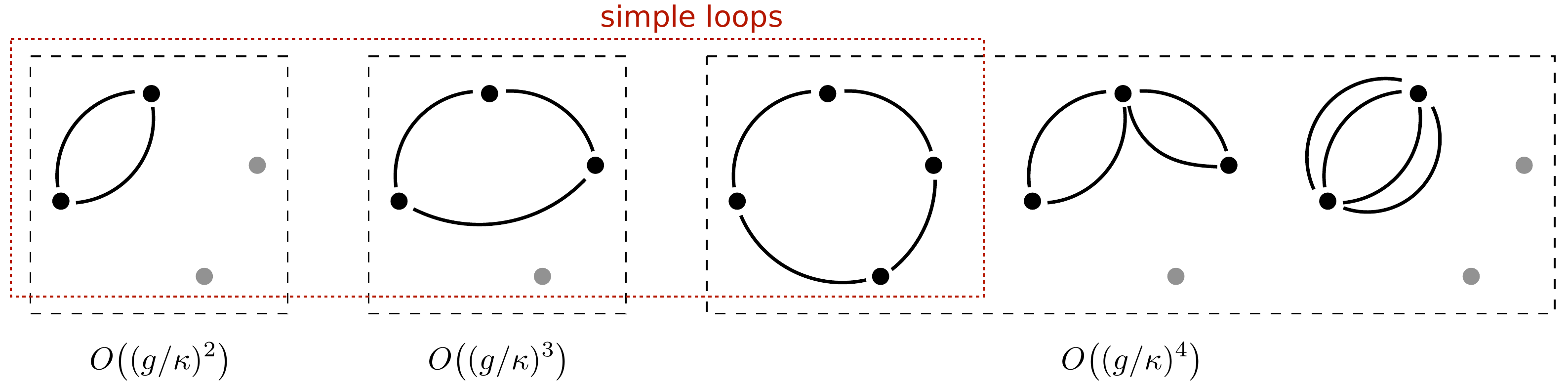}
    \caption{Allowed diagrams $\Lambda \in \mathcal{D}_\ell$ in the expansion \eqref{eq:diagram_expansion} up to order $\ell=4$. Including counting factors, they evaluate (from left to right) to $N\kappa/2$, $N\kappa(\kappa-1)/3$, $N\kappa(\kappa-1)(\kappa-2)/4$, $N \kappa (\kappa-1)/3$, and $N\kappa/8$. Circled in red are the one-loop, connected diagrams of which there are one per each order $\ell$: these contribute to lowest order in $1/\kappa$. Notice also that there is no watermelon diagram at $O((g / \kappa)^3)$ because of point (\ref{item:diagram_parity}) in the text.} 
    \label{fig:diagrams}
\end{figure}

To see it, start by expanding in powers of $g$ the full free energy density:
\begin{align}
    \nonumber
    -\beta f N &= \log \int [dq \, d\phi] e^{-\beta H_0 + \mu Q} \sum_{k=0}^\infty \frac{1}{k!} \bigg[ \frac{2g\beta}{\kappa} \sum_{ij} A_{ij} \sqrt{q_i q_j} \cos(\phi_i - \phi_j)\bigg]^k \\
    &=: -\beta f_0 N + \sum_{\ell=1}^\infty \left(\frac{2 g \beta }{\kappa} \right)^\ell \sum_{\Lambda \in \mathcal{D}_\ell} \Lambda
    \label{eq:diagram_expansion}
\end{align}
where consistently we denote by a ``0'' subscript quantities that are evaluated at $g=0$. Equation \eqref{eq:diagram_expansion} is our definition of the diagrams $\Lambda \in \mathcal{D}_\ell$, that we also show graphically in Fig.~\ref{fig:diagrams}. More precisely, at each order $\ell$ of the effective coupling constant $g / \kappa$ we have averages
\begin{equation*}
    \int [d\phi] \, \cos(\phi_{i_1} - \phi_{i_2}) \cos(\phi_{i_3} - \phi_{i_4}) \cdots \cos(\phi_{i_{2\ell-1}} - \phi_{i_{2\ell}}) \prod_{j=1}^N \langle q^{n_j/2} \rangle_{0} ,
\end{equation*}
where $n_j$ is the multiplicity with which index $j$ appears in the string $i_1 i_2 \cdots i_{2 \ell}$. We notice that
\begin{enumerate}
    \item since we are expanding a logarithm, by the linked-cluster theorem each diagram $\Lambda \in \mathcal{D}_\ell$ must consist of one connected piece only;

    \item \label{item:diagram_parity} for the angular integration not to yield 0, each $\phi_i$ must appear an even number of times; in particular this means that all the diagrams in $\mathcal{D}_\ell$ must be closed and each vertex must have an even number of legs (see Fig.~\ref{fig:diagrams}); 
    
    \item the permutation symmetry of the couples $i_{2p} i_{2p+1}$ yields a factor $\ell!/S_\Lambda$, where $S_\Lambda$ is the symmetry factor of the diagram $\Lambda$. Therefore, according to the usual arguments this cancels the $1/\ell!$ in the expansion of the exponential, leaving the symmetry factor in the denominator;
    
    \item the permutation symmetry within each couple $i_{2p} i_{2p+1}$ of the two indices yields a factor $2$ for each pair, and so a factor $2^\ell$ in total;
    
    \item \label{item:simple_loops} the angular integration for simple loops evaluates to $2^{1-\ell}$, while multiple loops give a result depending on the geometry (e.g.\ in Fig.~\ref{fig:diagrams} the first three diagrams are simple loops and receive respectively a factor $1/2$, $1/4$ and $1/8$, while the fourth receives a factor $1/4$ and the last a factor $3/8$).
    \medskip
\end{enumerate}

By using the previous rules, and having a look at Fig.~\ref{fig:diagrams}, one can get convinced that at fixed order $\ell$ the simple loops (e.g.\ the diagrams circled in red in Fig.~\ref{fig:diagrams}) are the least suppressed by $\kappa$. Indeed, since they are composed by the maximum number of \emph{distinct} points, the factor $\kappa^\ell$ in the denominator of Eq.~\eqref{eq:diagram_expansion} is compensated by the $\sim N \kappa^{\ell-1}$ possible choices of the points. Having noted this feature, we can explicitly compute $f_1$: the angular integration yields a factor $2^{1-\ell}$ (as noted in point (\ref{item:simple_loops}) before), the symmetry factors are $S_\Lambda = 2\ell$, and only the averages $\langle q^1 \rangle_0 = 1$ appear. Therefore one has
\begin{equation}
    \label{eq:f1}
    - \beta f_1= \frac{\kappa}{N} \sum_{\ell=2}^\infty \left( \frac{2 g \beta}{\kappa} \right)^\ell \, N \kappa^{\ell-1} \frac{1}{2\ell} 2^\ell \frac{1}{2^{\ell-1}}= - 2 g \beta - \log( 1 - 2 g \beta),
\end{equation}
with the sum starting from $\ell = 2$ because there is no diagram at order 1.

At this point, we can give a physical interpretation to Eqs.~\eqref{eq:TAP_expansion}--\eqref{eq:f1}. In the large connectivity limit $\kappa \to \infty$, the extensive contribution to the free energy is always regular and independent of the hopping between different sites. Moreover, as long as $\beta < \beta_s := (2g)^{-1}$, the sub-extensive contribution $f_1$ can be forgotten, while at $\beta = \beta_s$ it diverges and there is a phase transition: interactions must be taken into account and to go beyond one needs to address the problem non-perturbatively.

%------------------------------------------FULLY-CONNECTED------------------------------------------%  
\section{Solution of the fully-connected model}
\label{sec:fully_connected}

\begin{figure}[t]
    \centering
    \includegraphics[width=0.7\textwidth]{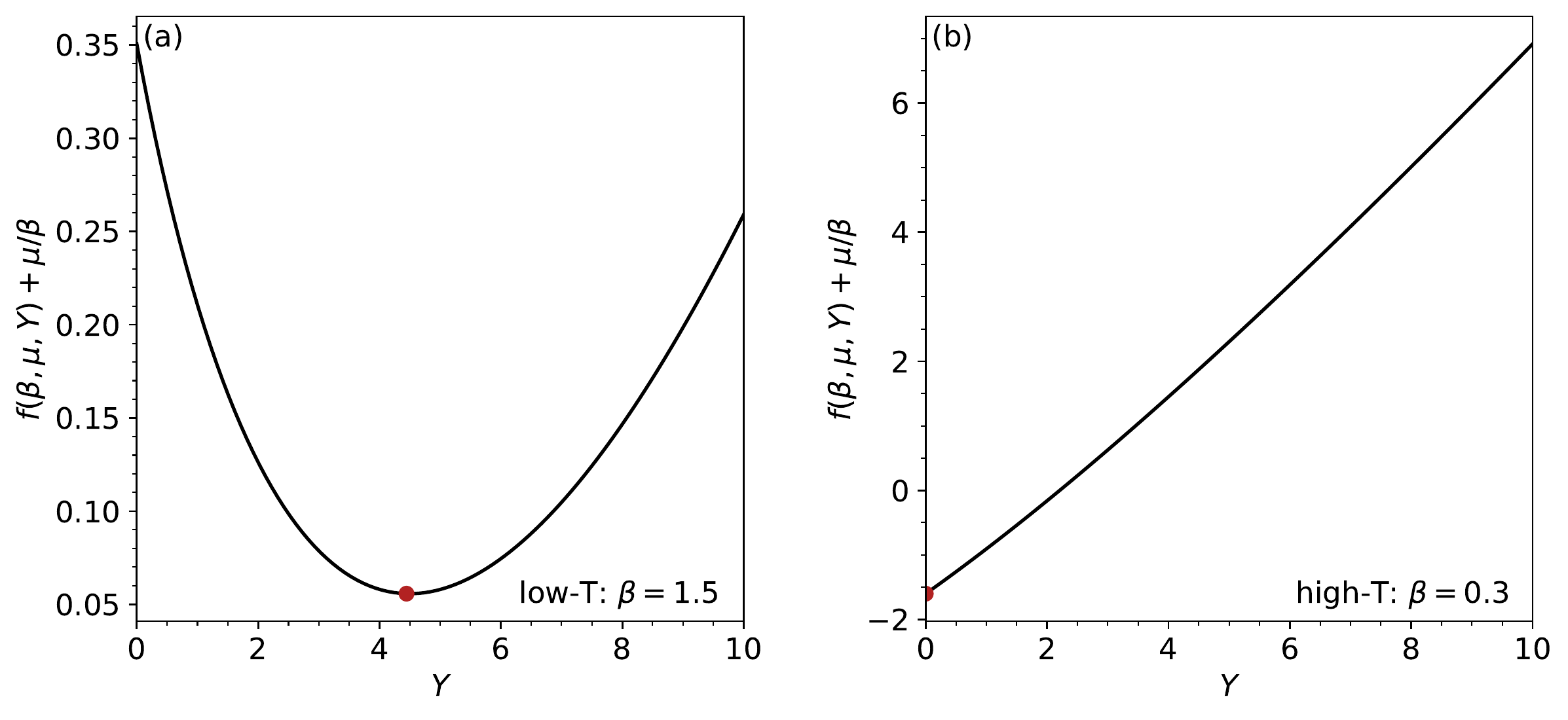}
    \caption{Plot of the free energy density $f(\beta,\mu,Y) + \mu/\beta$, with $g=1$, $v=2$, and $\mu$ fixed so that $\langle Q \rangle = N$. The red dot is the solution of Eq.~\eqref{eq:sp_Y}, i.e.\ the extremal point. It moves from the bulk of the allowed region $Y>0$ at low temperature (panel (a)), to the boundary $Y=0$ at high temperature (panel (b)).}
    \label{fig:fymu}
\end{figure}

To go beyond perturbation theory, we can compute the partition function of the fully-connected model $\kappa = N-1$ using saddle-point methods. Dropping sub-leading terms in $N$, we have
\begin{equation}
    Z_{\mathit{MF}}(N, \beta, \mu) = \int [dq\, d\phi] \exp \Big\{ -\frac{\beta v}{2} \sum_i q_i^2 + \mu \sum_i q_i + \frac{2 \beta g}{N} \Big(\sum_i \sqrt{q_i} e^{i\phi_i} \Big) \Big(\sum_i \sqrt{q_i} e^{-i\phi_i} \Big) \Big\}.
\end{equation}
We expand 
\begin{equation*}
    \Big(\sum_i \sqrt{q_i} e^{i\phi_i} \Big) \Big(\sum_i \sqrt{q_i} e^{-i\phi_i} \Big) =
    \Big( \sum_i \sqrt{q_i} \cos \phi_i \Big)^2 + \Big( \sum_i \sqrt{q_i} \sin \phi_i \Big)^2,
\end{equation*}
so that we can perform a Hubbard-Stratonovich transformation:
\begin{multline*}
    Z_{\mathit{MF}}(N, \beta, \mu) =  \frac{N}{2 \pi} \int [dq\, d\phi] \int dy_1 dy_2 \exp \Big\{ -\frac{\beta v}{2} \sum_i q_i^2 - \frac{N}{2} (y_1^2 +y_2^2)+ \mu \sum_i q_i \\
    + 2 \sqrt{\beta g} \, y_1 \sum_i \sqrt{q_i} \cos \phi_i + 2 \sqrt{\beta g} \, y_2 \sum_i \sqrt{q_i} \sin \phi_i \Big\}.
\end{multline*}
Now all the $q$,$\phi$ integrals are factorized, and the basics constituents are of the form
\begin{equation*}
    \frac{1}{2 \pi} \int dq\, d\phi \, \exp \Big[ -\beta v q^2/2 + \mu q + \sqrt{\beta g q} \, (y_- e^{i\phi} + y_+ e^{-i\phi} ) \Big]
\end{equation*}
with $y_{\pm} = y_1 \pm i y_2$. We can perform first the angular part:
\begin{align*}
    \frac{1}{2\pi}\int_0^{2 \pi} d\phi \, e^{z (y_- e^{i\phi} + y_+ e^{-i\phi} ) }
    &= \frac{1}{2\pi} \int_0^{2 \pi} d\phi \sum_{k \geq 0} \frac{z^k y_-^k}{k!} e^{i k \phi} \sum_{\ell \geq 0} \frac{z^\ell y_+^\ell}{\ell!} e^{-i \ell \phi} \\
    &= \sum_{k \geq 0} \frac{(z^2 y_+ y_-)^k}{(k!)^2}
    = I_0(2 z \sqrt{y_+ y_-}),
\end{align*}
$I_0$ being the modified Bessel function of the first kind. Thus, defining
\begin{equation}
    \label{eq:def_J}
    J(\beta,\mu,Y) := \int_0^\infty dq\, e^{-\beta v q^2/2 + \mu q} I_0 \big(2 \sqrt{\beta g q Y} \big)
\end{equation}
with $Y := y_+ y_- = y_1^2 + y_2^2$, we arrive at
\begin{equation}
    \label{eq:ZMFY}
    Z_{\mathit{MF}}(N, \beta, \mu) =  N \int_0^\infty dY e^{-NY/2 + N \log J(\beta, \mu, Y) }.
\end{equation}
When performing this integral in the $N\to\infty$ limit, if the saddle point is within the domain of integration $Y\geq 0$, one can use the saddle point method, otherwise one needs to integrate by parts around the lower limit of integration $Y=0$ (see Fig.~\ref{fig:fymu}). In any case, the free energy density is
\begin{equation}
    f(\beta,\mu,Y)=\beta^{-1}(Y/2-\log J),
\end{equation}
where $Y$ solves the saddle-point equation (with the above proviso)
\begin{equation}
    \label{eq:sp_Y}
    \frac{1}{2} = \frac{1}{J} \pd{J}{Y}.
\end{equation}
It also is convenient to trade $\mu$ for the (average) total charge $\langle Q \rangle =N$:
\begin{equation}
    \label{eq:tot_charge_constr}
    1 = \frac{1}{J} \pd{J}{\mu}.
\end{equation}
Equations \eqref{eq:sp_Y}--\eqref{eq:tot_charge_constr} can be easily solved numerically by iteration for any desired $\beta$.\medskip

Another way of rewriting Eqs.~\eqref{eq:sp_Y}--\eqref{eq:tot_charge_constr} is by interpreting $J(\beta,\mu,Y)$, defined in Eq.~\eqref{eq:def_J}, as a partition function for the variable $q$, which thus acquires the probability density
\begin{equation}
    \label{eq:p(q)}
    p(q) = \frac{1}{J} e^{-\beta v q^2/2 + \mu q} I_0 \big(2 \sqrt{\beta g q Y}\big).
\end{equation}
Then, the two equations \eqref{eq:sp_Y}--\eqref{eq:tot_charge_constr} take (respectively) the form
\begin{align}
    \label{eq:sp_Y_probdistr}
    \sqrt{\frac{Y}{4\beta g}} & =\left\langle \sqrt{q} \, \frac{I_1(2 \sqrt{\beta g q Y})} {I_0(2 \sqrt{\beta g q Y})} \right\rangle_p\\
    \label{eq:q=1}
    1 &= \left\langle q \right\rangle_p.
\end{align}

These last expressions are convenient to control the limits $\beta \to \infty$ and $\beta \to 0$. Indeed, as $\beta \to \infty$ the problem simplifies and the probability concentrates around the saddle point $q=1$ (Eq.~\eqref{eq:q=1}). One can also expand the Bessel functions (as long as, self-consistently, $Y\gg 1/\beta$) for large arguments, and substituting $q=1$ in Eq.~\eqref{eq:sp_Y_probdistr} gives
\begin{equation}
    \label{eq:Y_solution_betaInfty}
    \sqrt{\frac{Y}{4\beta g}} = \langle \sqrt{q} (1+ \cdots) \rangle_p \implies Y = 4 g \beta + O(\beta^0).
\end{equation}
Also, imposing Eq.~\eqref{eq:q=1} explicitly on Eq.~\eqref{eq:p(q)}, one gets
\begin{equation}
    \label{eq:mu_solution_betaInfty}
    \mu = \beta (v - 2g) + O(\beta^0).
\end{equation}

\begin{figure}[t]
    \centering
    \includegraphics[width=0.45\textwidth]{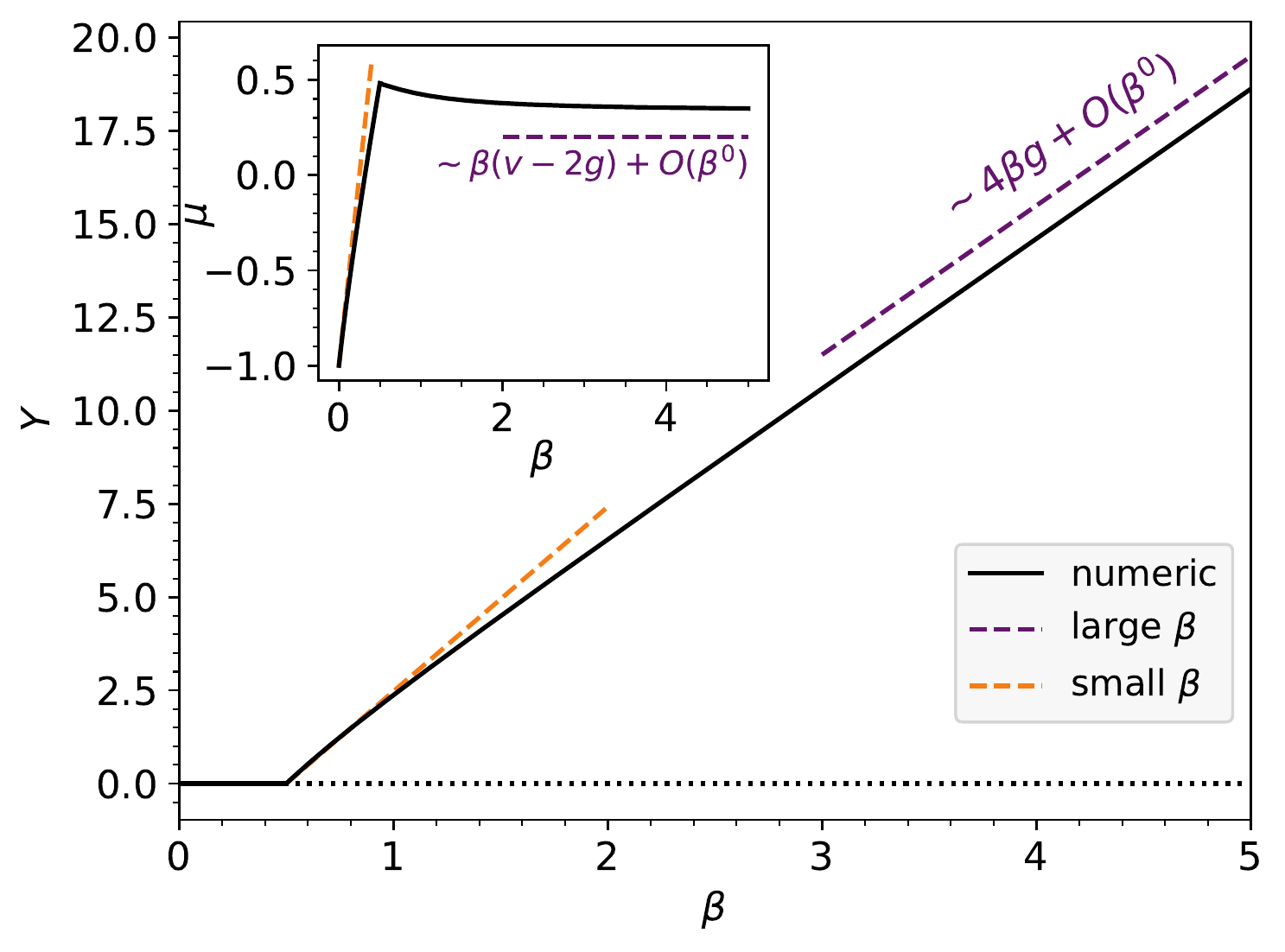}
    \caption{Saddle-point value of $Y$ found upon solving Eqs.~\eqref{eq:sp_Y}--\eqref{eq:tot_charge_constr} by iteration, with $g=1$ and $v=2$ (black solid line). For $\beta<(2g)^{-1}=0.5$ the correct solution is $Y_1$, while at larger values of $\beta$ it becomes $Y_2$ (see Eq.~\eqref{eq:Y_solutions_beta0}). For comparison, we show the approximate solutions at $\beta \to \infty$ and $\beta Y \to 0$ as dashed lines. Note that the $\beta \to \infty$ approximation, to the order obtained in Eqs.~\eqref{eq:Y_solution_betaInfty}--\eqref{eq:mu_solution_betaInfty}, still needs a $O(\beta^0)$ term to be fixed. \textit{Inset:} Corresponding values found for $\mu$. }
    \label{fig:Ysolution}
\end{figure}

For small $\beta$, instead, one can expand the Bessel functions for small argument (as long as this returns self-consistently $\beta Y\ll 1$), and obtain
\begin{align*}
    \sqrt{\frac{Y}{4\beta g}} 
    &= \left\langle \sqrt{q} \left[ (\beta g q Y)^{1/2} - \frac{1}{2} (\beta g q Y)^{3/2} + \cdots \right] \right\rangle_p \\
    &= (\beta g Y)^{1/2} \langle q\rangle_p - \frac{1}{2} (\beta g Y)^{3/2} \langle q^{2} \rangle_p+ \cdots \ .
\end{align*}
There are two solutions:
\begin{equation}
    \label{eq:Y_solutions_beta0}
    Y_1 = 0, \qquad
    Y_2 = \frac{2\beta g-1}{(\beta g)^2 \langle q^{2} \rangle_p}.
\end{equation}
The second solution is negative for $\beta < \beta_s = (2g)^{-1}$, so in this region one must stick with $Y_1$ (since the $Y$ integral in Eq.~\eqref{eq:ZMFY} is on the positive domain). As $\beta \gtrsim \beta_s$, instead, $Y_2$ becomes the correct solution, until the condition $\beta Y\ll 1$ is no more valid and the approximation breaks down. In Fig.~\ref{fig:Ysolution} we show the comparison of the numerically exact solutions with the small-$\beta$ and large-$\beta$ approximations.  \medskip

To connect with the diagrammatic expansion done in Sec.\ \ref{sec:large_connectivity}, we notice that the critical value $\beta_s = (2g)^{-1}$ is the same given by the radius of convergence of perturbation theory for $f_1$, the sub-extensive contribution to the free energy. We are now in position to give an interpretation to the phase transition taking place at $T_s = 1/\beta_s$: it is the temperature below which the angles $\phi_i$ no more average to zero, but start acquiring a common orientation. Indeed, on one hand
\begin{equation*}
    - g \pd{}{g} (\beta f)=\beta \frac{2g}{N^2}\sum_{i \neq j} \big\langle \sqrt{q_i q_j} \cos(\phi_i - \phi_j) \big\rangle;
\end{equation*}
on the other hand, by differentiating the saddle-point free energy,
\begin{equation*}
    - g \pd{}{g} (\beta f)
    = g \frac{1}{J} \pd{J}{g} 
    = Y \frac{1}{J} \pd{J}{Y} = \frac{Y}{2}.
\end{equation*}
The comparison of the last two equations implies
\begin{equation}
    \label{eq:meaning_Y}
    Y = \frac{4 \beta g}{N^2} \sum_{i \neq j} \big\langle \sqrt{q_i q_j} \cos(\phi_i - \phi_j) \big\rangle .
\end{equation}
We conclude that, as $\beta \to \infty$, the angles must all point in the same direction (albeit the latter can change in time). Indeed, recalling that $q$ concentrates around $1$, in order to find the asymptotic, low temperature behaviour $Y \simeq 4 \beta g$ one needs that all the phases align: $\langle \cos(\phi_i-\phi_j)\rangle\to 1$, so $\phi_i \to \phi_0$ for all $i=1,2,\dots,N$. This is the statistical mechanics signature of a synchronized phase \cite{Kuramoto1975Self,Acebron2005RMP}, in which all fields have a common phase and the fluctuations of the amplitudes are negligible. The synchronization phase transition is second-order, with the order parameter $Y$ growing linearly close to $\beta_{s}=(2g)^{-1}$.

We can also express the above observations in terms of the energy density $\epsilon$. At $T=0$ the system is in the ground state, with energy density $\epsilon_{\mathit{GS}} = v/2 - 2g$: this readily follows from our $\beta \to \infty$ expansion of the free energy. At the synchronization transition $T=T_s=2g$, instead, the energy density can be found numerically by imposing $Y=0$ and fixing $\mu$ from Eq.~\eqref{eq:tot_charge_constr}: for $g=1$ and $v=2$ we find $\epsilon_s = 1.481\dots$ (see also Fig.~\ref{fig:phase_diagram}).

Finally, we can identify the order parameter $Y$ with the average interaction energy density (see Eq.~\eqref{eq:meaning_Y}), which vanishes at temperatures $T > T_s$.

%------------------------------------------ANALYTICS------------------------------------------%  

\section{Topological structure of the potential energy surface}
\label{sec:topology}

We now start focusing on the region $\epsilon \geq \epsilon_c = v$. Having completely lost the spatial structure given by the hopping \emph{for any graph geometry}, the model has become \emph{effectively non-interacting}. For this reason, we can also fix $v \equiv 2$ wlog.\ from now on.

The microcanonical surface is non-trivial, because of the presence of two conservation laws: energy ($H=N \epsilon$) and charge ($Q=N$). For large energy density, the bulk of the volume of the microcanonical surface is concentrated in the region where a few charges get a large share of the total charge (the participation ratio is $O(1)$ \cite{Gradenigo2021Localization,Gradenigo2021Condensation}). These are localized charge configurations. However, these configurations are not isolated from each other, and a continuous charge rearrangement can move any localized lump anywhere else in space, passing through regions of equally distributed charges. In this section we show that this can be done by moving continuously on the microcanonic surface for any $\epsilon=O(1)$.

\begin{figure}[t]
    \centering
    \includegraphics[width=0.7\columnwidth]{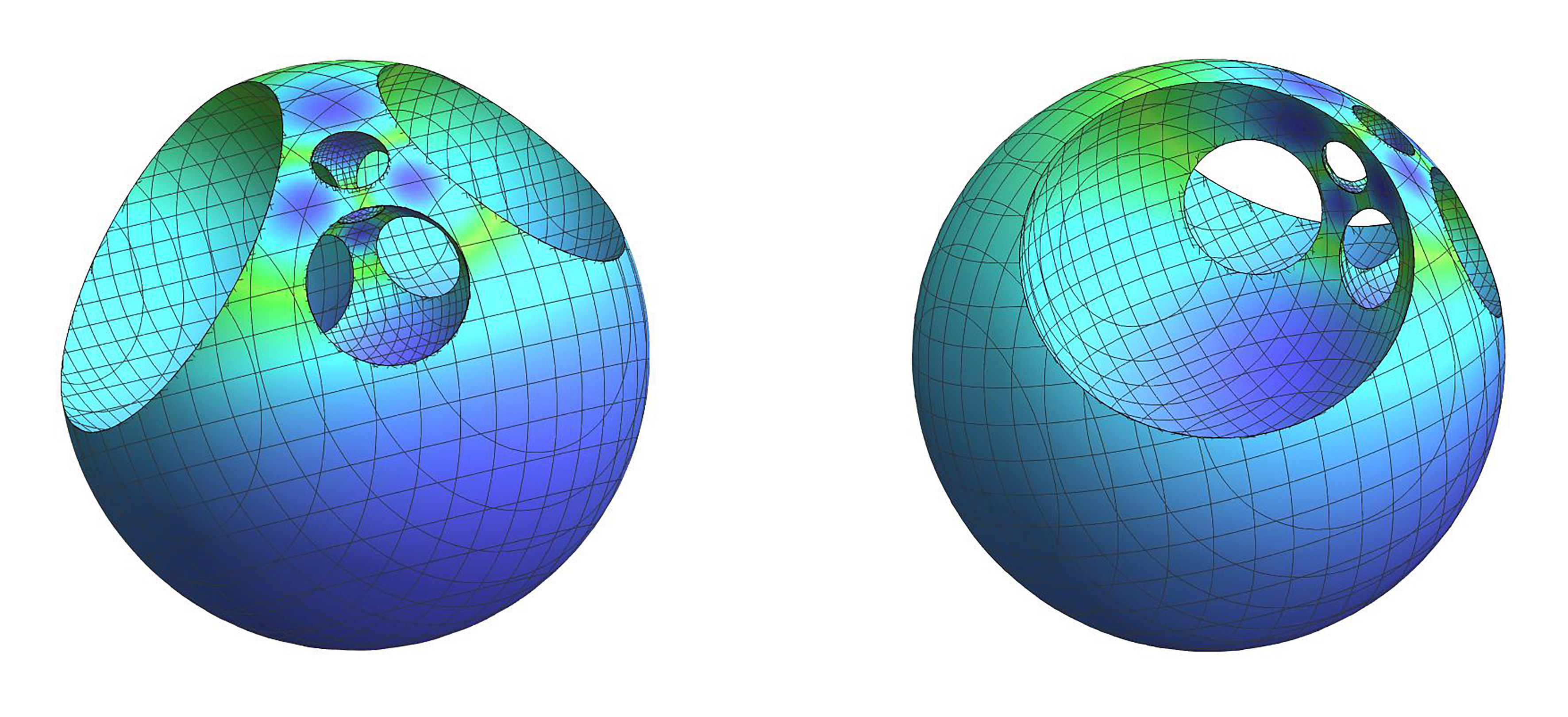}
    \caption{Two different views of the same stereographic projection of the manifold $\mathcal{M}_\epsilon$  (Eq.~\eqref{eq:manifold}), for $N=5$ and $\epsilon=2$. While this projection respects the topology of the manifold, clearly its metric structure is altered. Colors depend on the distance from the vertices of $\Delta^{N-1}$, varying from blue for the five 0-handles around those vertices to green for the median sections of the ten 1-handles.}
    \label{fig:manifoldN5}
\end{figure}

Let us summarize here what is proven as a theorem in this Section:
\emph{The microcanonical surface remains a connected manifold for all the energy densities $\epsilon<N/2$ (extensive energy density), in particular through the dynamical transition observed numerically at $\epsilon_c=2$, which then cannot be addressed to a deficiency in connectivity.} Moreover, we show that, as $\epsilon$ increases, the surface passes a series of critical points, according to stratified Morse theory. \emph{At the dynamical transition $\epsilon_c=2$, the number of transverse dimensions of the pipe connecting two regions corresponding to localized charges equals the number of longitudinal dimensions.} What this topological proof cannot tell us is \emph{how} the equilibration time depends on $N$. That is a property of the dynamics, which we can only conjecture is due to the shape of the pipes linking the ``fat" regions of localized charge, and it is presented at the end of this Section as a \emph{Problem}.

In order to prove these two results we will introduce now the manifold and some relevant results of Morse theory. We again change variables from the $\psi_i$'s to the local charges $q_i = |\psi_i|^2$ (see also Eq.~\eqref{eq:psi_to_qphi}), that are the only combinations of the $\psi_i$'s entering in the conservation laws. Thus, we are left with the equations
\begin{equation}
    \label{eq:manifold}
    \begin{cases}
        \frac{1}{N} \sum_{i=1}^N q_i = 1 \\
        \frac{1}{N} \sum_{i=1}^N q_i^2 = \epsilon \\
        q_i \geq 0 \quad \forall i=1,2,\dots,N.
    \end{cases}
\end{equation}
These equations define a $(N-2)$-dimensional manifold with boundary and corners $\mathcal{M}_\epsilon$ (see Fig.~\ref{fig:manifoldN5} for a visual impression of the case $N=5$), naturally embedded in $\mathbb{R}^N$, whose central role was recognized already in \cite{Chatterjee2017Note}. The topology of this manifold undergoes a series of changes as $\epsilon$ varies, which can be outlined by stratified Morse theory in the following way (see App.\ \ref{app:topology} for a more detailed description).

The first and the last equation in \eqref{eq:manifold} represent the affine simplex $\Delta^{N-1} \subset \mathbb{R}^N$ spanned by the vectors $Ne_1, \dots, Ne_N$, where $e_1,\dots, e_N$ is the canonical base of $\mathbb{R}^N$. Hence, $\mathcal{M}_\epsilon$ is non-empty for $1 \leq \epsilon \leq N$. Moreover, $\mathcal{M}_\epsilon$ is a small $(N-2)$-sphere around the barycenter of $\Delta^N$ when $\epsilon$ approaches 1, while it is the disjoint union of $N$ small $(N-2)$-disks, each near to a vertex of $\Delta^{N-1}$, when $\epsilon$ approaches $N$. 

In order to see what happens for the intermediate values of $\epsilon$, think of the boundary $\partial \Delta^{N-1}$ as a stratified space, whose strata are its open sub-simplices of $\Delta^{N-1}$, and observe that $\varphi:\partial \,\Delta^{N-1} \to \mathbb{R}$ given by $\varphi(q) = \|q\|^2/N$ is a stratified Morse function, meaning that it restricts to a Morse function on every stratum. Then, for every $1 < \epsilon < N$, the radial projection from the barycenter of $\Delta^{N-1}$, that is the vector $e_1 + \dots + e_N$, induces a stratified diffeomorphism between $\mathcal{M}_\epsilon$ and the suplevel set $M^\epsilon(\varphi) = \{q \in \partial \Delta^{N-1} \;|\; \varphi(q) \geq \epsilon\} \subset \partial \Delta^{N-1}$, according to the second equation in \eqref{eq:manifold}. 

Morse theory tells us that the topology of $\mathcal{M}_\epsilon \cong M^\epsilon(\varphi)$ changes only at the critical values of the restrictions of $\varphi$ to the strata of $\partial \Delta^{N-1}$. Such critical values have the form $N/k$ with $1 < k < N$. Indeed, for each $k$ we have $\binom{N}{k}$ corresponding non-degenerate critical points of index $N-k-1$ located at the barycenters of the $(k-1)$-dimensional faces of $\Delta^{N-1}$. This implies that for $\delta > 0$ small enough $M^{N/k \,-\, \delta}(\varphi)$ can be obtained by attaching $\binom{N}{k}$ narrow $(k-1)$-handles to $M^{N/k \,+\, \delta}(\varphi)$. Each of these $(k-1)$-handles is an $(N-2)$-cell $C^{N-2}$ which is the product of a $(k-1)$-cell $C^{k-1} = \mathrm{Cl}(\Sigma - M^{N/k \,+\, \delta}(\varphi))$ (where $\mathrm{Cl}$ stands for the closure operator)  for a $(k-1)$-dimensional face $\Sigma$ of $\Delta^{N-1}$ and a small $(N-k-1)$-cell $C^{N-k-1}$ such that $C^{N-2} \cap M^{N/k \,+\, \delta}(\varphi) = \partial \,C^{k-1} \times C^{N-k-1}$.

\begin{figure}[t]
    \centering
    \includegraphics{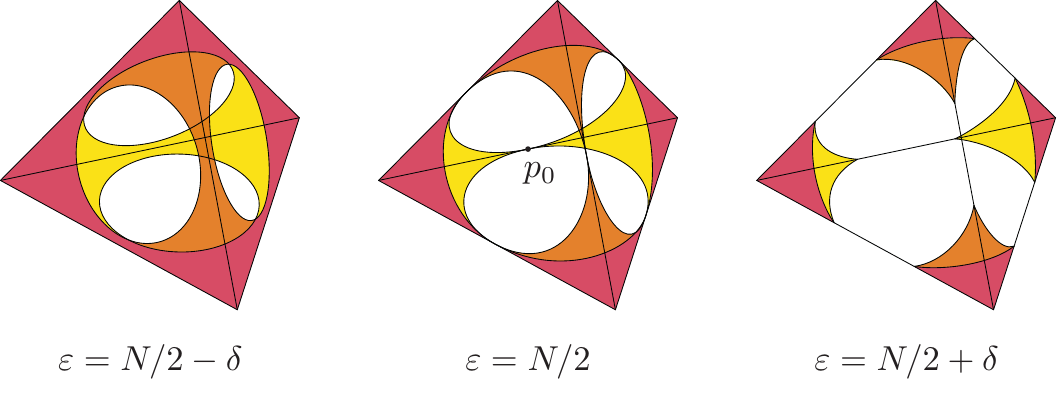}
    \caption{The last critical point, at which the microcanonic manifold splits into $N$ disconnected pieces each representing a different set of localized configurations, close to each of the vertices of $\Delta^{N-1}$. The figure concerns the case $N = 4$. Everything is depicted in the $(N-1)$-dimensional affine subspace $A^{N-1}$ given by the first equation in \eqref{eq:manifold}. In yellow $\mathcal M_\epsilon \subset \mathbb{S}^{N-2}_\epsilon$ and in red the suplevel set $M^\epsilon(\varphi) \subset \mathrm{Bd}\,\Delta^{N-1}$.}     \label{fig:manifoldN4}
\end{figure}

As a consequence, for every $k = 1,\dots,N-1$ and $N/(k+1) < \epsilon < N/k$ the $(N-2)$-manifold $M^\epsilon(\varphi)$ is a regular neighborhood, meaning an $(N-2)$-dimensional thickening, of the $(k-1)$-skeleton of $\Delta^{N-1}$ in $\partial \Delta^{N-1}$. In particular, recalling the homeomorphism $\mathcal{M}_\epsilon \cong M^\epsilon(\varphi)$, we can conclude that $\mathcal{M}_\epsilon$ has $N$ connected components for $N/2 < \epsilon \leq N$, while it is connected for $1 \leq \epsilon \leq N/2$.

The above discussion shows that, for $N>4$, at $\epsilon =2$ the manifold $\mathcal{M}_\epsilon$ has gone through a series of gluing handles procedures described above, yet remaining connected. This raises the intriguing problem of characterizing $\mathcal{M}_2$ also from a purely
geometrical point of view. Of course the simplest geometrical invariant of $\mathcal{M}_\epsilon$ is its volume. Since $\mathrm{vol}(\mathcal{M}_1)=\mathrm{vol}(\mathcal{M}_N)=0$, we know by continuity that there exists $\epsilon_0(N) \in (1,N)$ which maximizes $\mathrm{vol}(\mathcal{M}_\epsilon)$. Recall that Boltzmann's law entails $S(\epsilon)=\log(\mathrm{vol(\mathcal{M}_\epsilon)})$, and also that it holds 
\begin{equation*}
    \frac{1}{T} = \frac{1}{N} \der{S}{\epsilon}.
\end{equation*}
At infinite temperature clearly $\der{S}{\epsilon}|_{\epsilon=2}=0$. So, the stationary point for the microcanonical manifold volume arises at $\epsilon = 2$ (value that is correct only in the limit $N\to \infty$), as already argued before. This classical observation has been significantly strengthened in \cite{Gradenigo2021Localization,Gradenigo2021Condensation} for the model under consideration, where it is observed that $\epsilon = 2$ is indeed $\lim_{N\to \infty}\epsilon_0(N)$ and moreover $\epsilon_0(N) = 2 + O(N^{-1/3})$. We believe that a direct geometric analysis of the behaviour of $\mathrm{vol}(\mathcal{M}_\epsilon)$ would be very interesting by itself since it could shed light on various other aspects of the problem studied.

\begin{figure}[t]
    \centering
    \includegraphics[width=0.7\columnwidth]{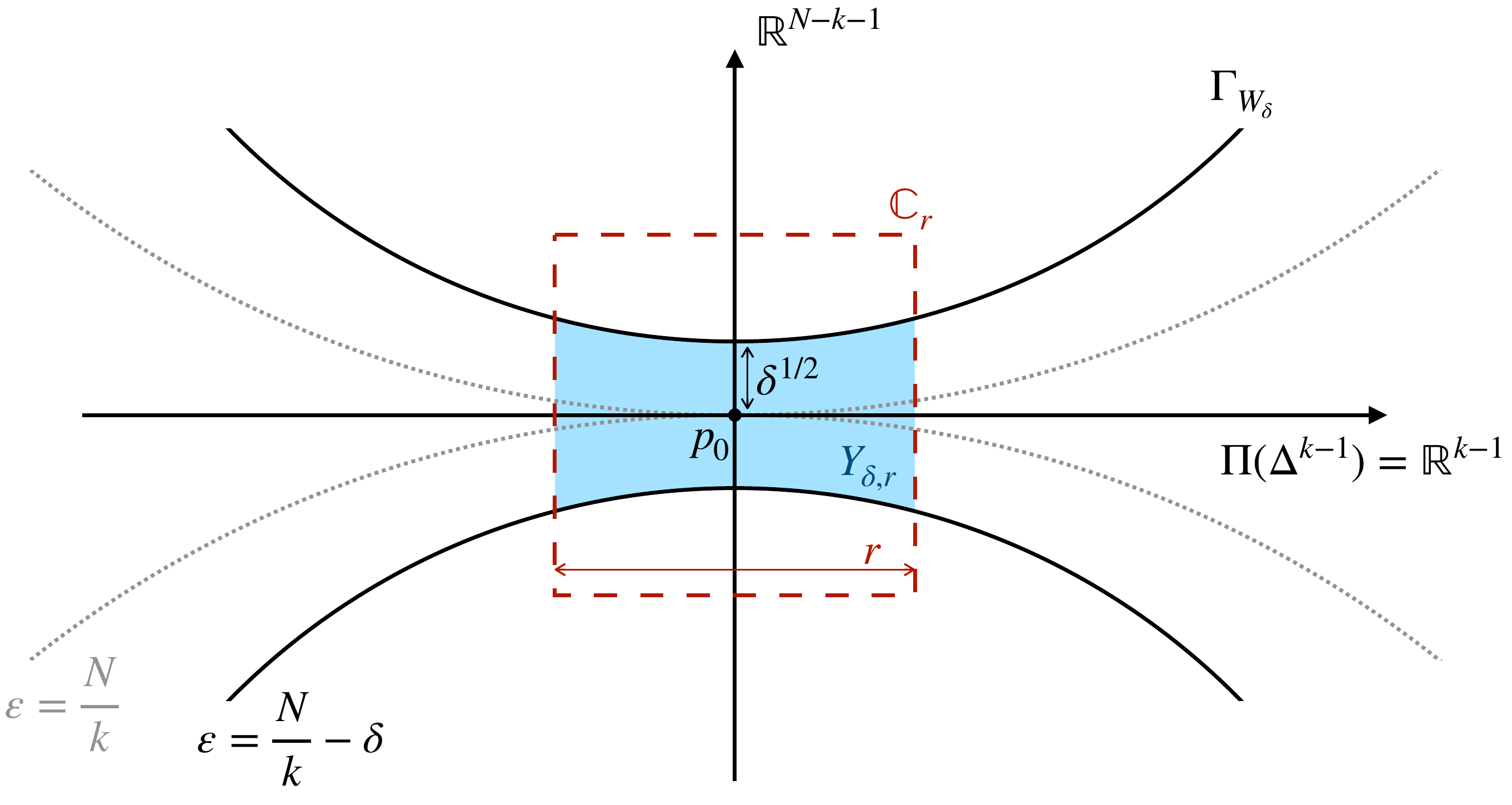}
    \caption{Image of the projection via $\Pi$ on $T_{p_0}(\mathbb{S}^{N-2})$ of a neighborhood of a singular point $p_0$ in $\mathcal{M}_{N/k}$. $\Gamma_{W_\delta}$ is the profile of $\Pi(\mathcal{M}_{N/k - \delta})$ in $T_{p_0}(\mathbb{S}^{N-2})$. In blue the handle attachment $Y_{\delta,r}$.}
    \label{fig:projection}
\end{figure}

While we leave this task for future investigation, we now observe that thanks to the Morse-theoretic description above, we can quantify the volume contribution of each handle attachment through any critical value of $\epsilon =N/k$, $1<k<N$. Indeed, given $\epsilon = N/k - \delta$ and $p_0$ a singular point in $\mathcal{M}_{N/k}$, we can look at the projection $\Pi$ from the barycenter $B$ of the symplex of a neighborhood of $p_0$ in the sphere $\mathbb{S}^{N-2}$ inside the $(N-1)$-dimensional affine subspace $A^{N-1}$ given by the first equation in \eqref{eq:manifold} onto the tangent space to this sphere (see Figs.\ \ref{fig:manifoldN4} and \ref{fig:projection}).

As argued above, $p_0$ is a non-degenerate critical point of index $N-k-1$ located at the barycenter of a $(k-1)$-dimensional face $\Delta^{k-1}$ of $\Delta^{N-1}$ and hence we can choose coordinates $(x_1, \dots, x_{k-1}, y_1, \dots, y_{N-k-1})$ on $T_{p_0}(\mathbb{S}^{N-2})$ in such a way that $\underline{x}=(x_1, \dots, x_{k-1})$ parametrize $\Pi(\Delta^{k-1})$, and $\underline{y}=(y_1, \dots, y_{N-k-1})$ span its orthogonal complement. By intersecting $\Pi(\mathcal{M}_\epsilon)$ with a $(N-2)$-cube $\mathbb{C}_r$ centered in the origin of $T_{p_0}(\mathbb{S}^{N-2})$ with faces parallel to the coordinate axis, we are led to estimate the rate change of the local effect on the volume of the handle-attachment procedure ($\mathrm{vol}(Y_{\delta,r})$ as shown in Fig. \ref{fig:projection}). This can be done observing that such region is bounded by a function $W_\delta(|\underline{x}|)$, which is at first order quadratic in $|\underline{x}|$, being the image via $\Pi$ of the profile of the sphere, and s.t. $W_\delta(0)=\delta^{1/2} + O(\delta)$. It is now a straightforward computation to see that 
\begin{equation}
    \label{eq:volume_estimate}
    \mathrm{vol}(Y_{\delta,r}) = C(N,k) \, r^{k-1} \delta^{(N-k-1)/2} + \mathrm{h.o.}   
\end{equation}
for some constant $C(N,k)$. The above computation holds for any $k = 2 \dots N-1$, and singles out yet another peculiarity of the value $\epsilon = 2$, corresponding to $k= N / 2$ (for even values of $N$). In fact, this is the only situation in which the contributions coming from the two factors of the handle $C^{k-1}\times C^{N-k-1}$ are of the same order. \medskip

Having established that nothing worth of notice in the topology of $\mathcal{M}_\epsilon$ occurs at $\epsilon=2$, we will see in the next Section that a simple Brownian motion on $\mathcal{M}_\epsilon$ does change its behavior precisely at $\epsilon=2$. The dynamics of the Brownian motion is notoriously linked to another natural geometric invariant of $\mathcal{M}_\epsilon$, namely its first non-zero eigenvalue of the Laplacian for the curved metric induced on $\mathcal{M}_\epsilon$ (with Neumann boundary conditions). 

We will then provide in the next Section strong evidence for the following intriguing (and hard) geometric \\ \hspace{\textwidth}
\textbf{Problem.} \hspace{0.1cm} \emph{Having set $\gamma(\epsilon) := - \lim_{N\to \infty} \frac{1}{N} \log \lambda_1$, we have}
\begin{equation*}
\begin{cases}
    \gamma  = 0 &  \epsilon \leq 2 \\  
    \gamma  >0  &  \epsilon >2.
\end{cases}
\end{equation*}

Providing fine estimates for the first eigenvalue of the Laplacian is well known to be a subtle (and important) problem in geometric analysis.
The present situation seems particularly interesting and original also from a purely mathematical point of view for the concurrence of the value
$\epsilon = 2$ as special value both for the volume and $\lambda_1$, a coincidence that certainly deserves further understanding on the mathematical side. \medskip

We believe, however, that the simple observation in Eq.~\eqref{eq:volume_estimate} could be a first step towards the understanding of the coincidence stated above. Indeed, a very much conjectural, and simplified picture of why the charges become localized could be based on the counting of ``useful" and ``useless" directions when crossing the handles connecting two different localized configurations. One can make as well a connection with the question of \emph{entropic barriers} in spin-glass dynamics (on this topic see e.g.\ \cite{krzakala2007gibbs,Auffinger2013Complexity,Auffinger2013Random,bellitti2021entropic}). We leave this connection for future investigations.

%------------------------------------------NUMERICS------------------------------------------%  

\section{A Brownian dynamics on the potential energy surface and the gap of the Laplacian}
\label{sec:dynamics}

In order to extract the first non-zero eigenvalue of the Laplacian, we resorted to studying the correlation functions of a Brownian motion on the surface $\mathcal{M}_\epsilon$. Indeed, being the diffusion equation described by the Laplacian, it is known that the late decay of the correlation functions of coordinates (e.g.\ the charges $q_i$) gives its first non-zero eigenvalue. Hence, we pick as a starting point a random vector $\vec q$ that satisfies all the conditions in \eqref{eq:manifold} (this can be easily done by repeatedly projecting on the three distinct manifolds defined by each constraint, until they are all obeyed), and let it evolve by free diffusion on $\mathcal{M}_\epsilon$ up to a final time $T_f$. Specifically, at each Monte Carlo step we update the position as $\vec q(t+dt) = \vec q(t) + d\vec{W}$, where $dW_i$ are i.i.d.\ Gaussian random variables s.t.\ $\langle dW_i \rangle =0$ and $\langle dW_i^2 \rangle =dt$, $dt$ being small\footnote{Notice that with this normalization $\|\vec q(t+dt)- \vec q(t)\|=O(\sqrt{N})\sqrt{dt}$ and the relative Fokker-Plank equation is Eq.~\eqref{eq:FP}, which does not contain any explicit factor of $N$. Different scalings of $dq$ can be easily obtained by rescaling time.}; and then we enforce again the constraints until they are all satisfied (see App.\ \ref{app:numerics} for more details).

We believe it is important to emphasize that our dynamics is fundamentally different from that of \cite{Iubini2013Discrete,Iubini2014Coarsening,Gradenigo2021Localization,gotti2020finitesize}. In these works, the basic Monte Carlo step was the redistribution of charge within a triplet of sites. Specifically, a triplet $(q_{i}, q_{j}, q_{k})$ was updated to a randomly chosen new triplet $(q'_{i}, q'_{j}, q'_{k})$, with the constraint that the transformation $(q_{i}, q_{j}, q_{k}) \longmapsto (q'_{i}, q'_{j}, q'_{k})$ could be performed continuously in the subsystem defined by the three charges only, and without violating the (local) charge and energy constraints. In the case of \emph{consecutive} triplets $ i= j-1 = k-2$, this Monte Carlo algorithm provides a good description for the dynamics of a chain. The case of generic $i,j,k$, instead, addresses a mean-field situation like the one considered in this work. We believe nevertheless that it would be difficult to connect this ``triplet" dynamics to the Brownian motion (which instead is related to the eigenvalues of the Laplacian), so we decided to simulate directly the latter. \medskip

\begin{figure}[t]
    \centering
    \subfloat[]{\includegraphics[width=0.45\textwidth]{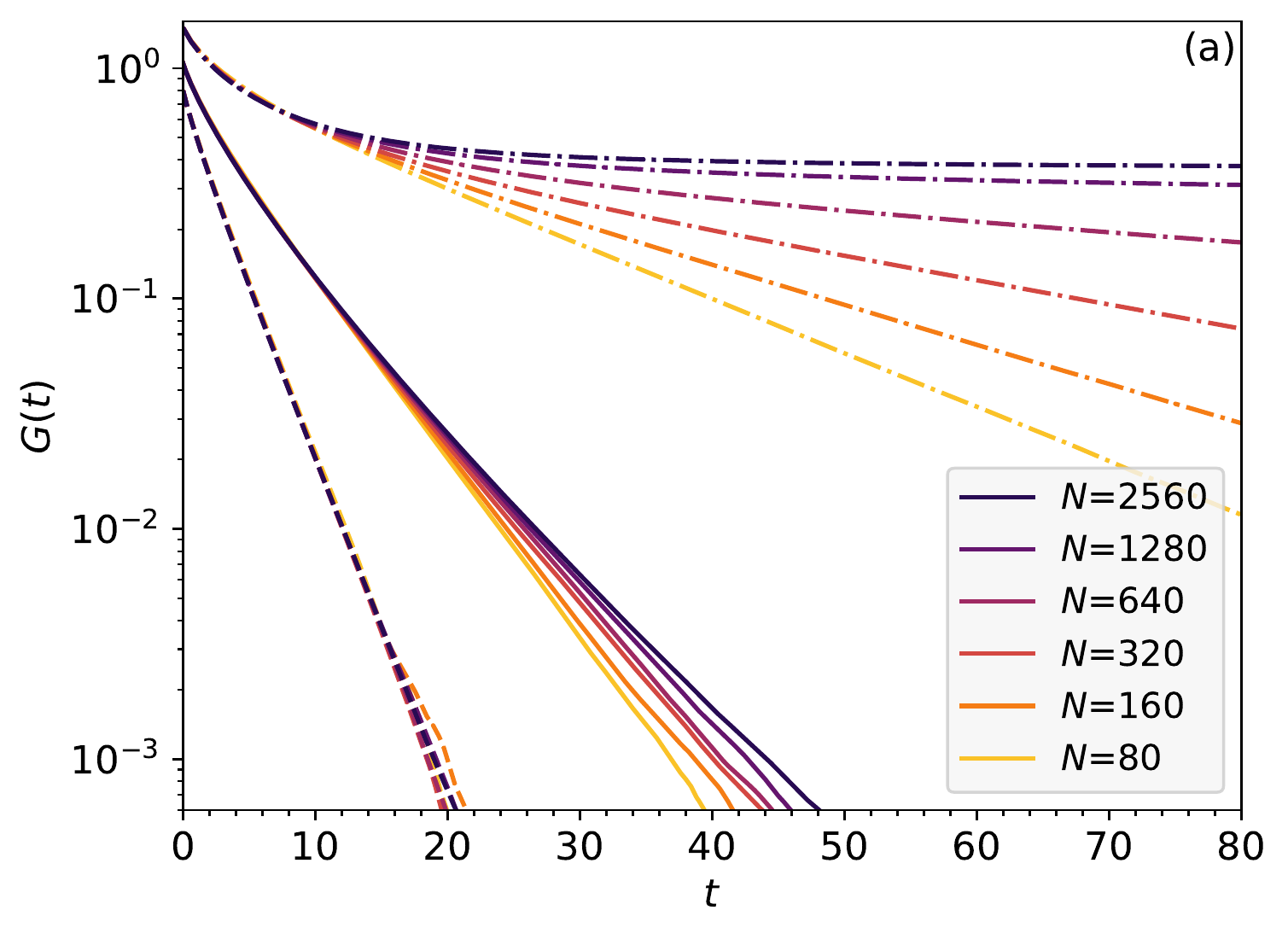}
    \label{fig:Gq_nonCritical}}
    \subfloat[]{\includegraphics[width=0.45\textwidth]{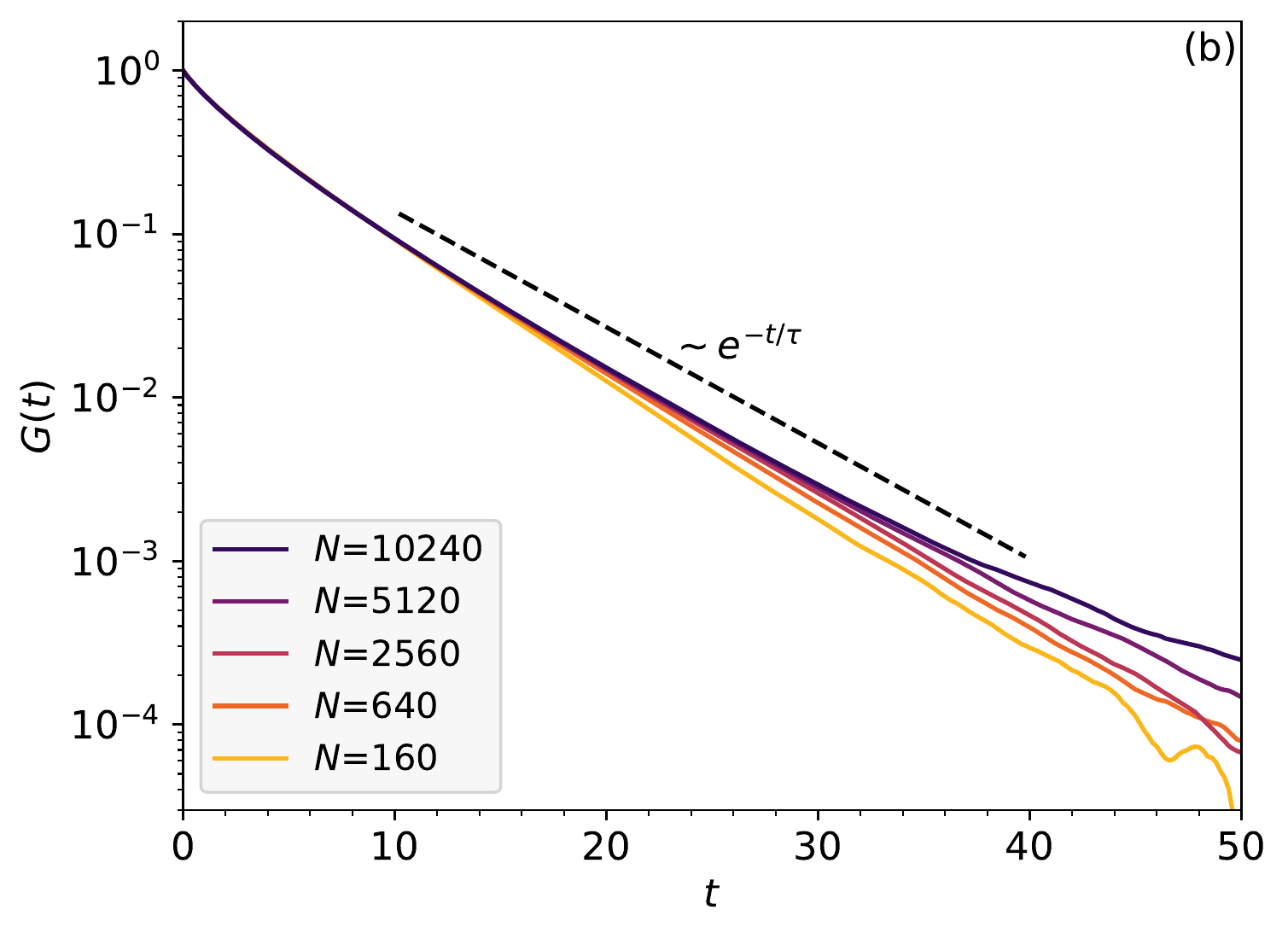}
    \label{fig:Gq_critical}}
    \caption{(a) Correlation function, Eq.~\eqref{eq:Gq}, as a function of physical time $t$ with $dt = 0.01$. Dashed lines refer to $\epsilon=1.9$ (ergodic region), solid lines to $\epsilon = 2.05$ (near-critical region) and dashed-dotted lines to $\epsilon = 2.5$ (localized region). One can see that, except in the critical region, the decay has a wide simple exponential window. Each curve is obtained by averaging over at least 5000 different runs. \\\hspace{\textwidth}
    (b) Correlation function at the critical point $\epsilon = 2$. The decay is slower than an exponential (and becomes slower as $N$ is increased), as shows the comparison with the black dashed line. For each $N$ we performed a fit $\log G(t) = -t /\tau - \log (1 + (t/t_1)^z)$, finding values of $\tau$, $t_1$ and $z$ that we report in Fig.~\ref{fig:fit_critical}. Each curve is obtained by averaging over at least 50000 different runs. }
    \label{fig:Gq}
\end{figure}

\begin{figure}[t]
    \centering
    \subfloat[]{\includegraphics[width=0.45\textwidth]{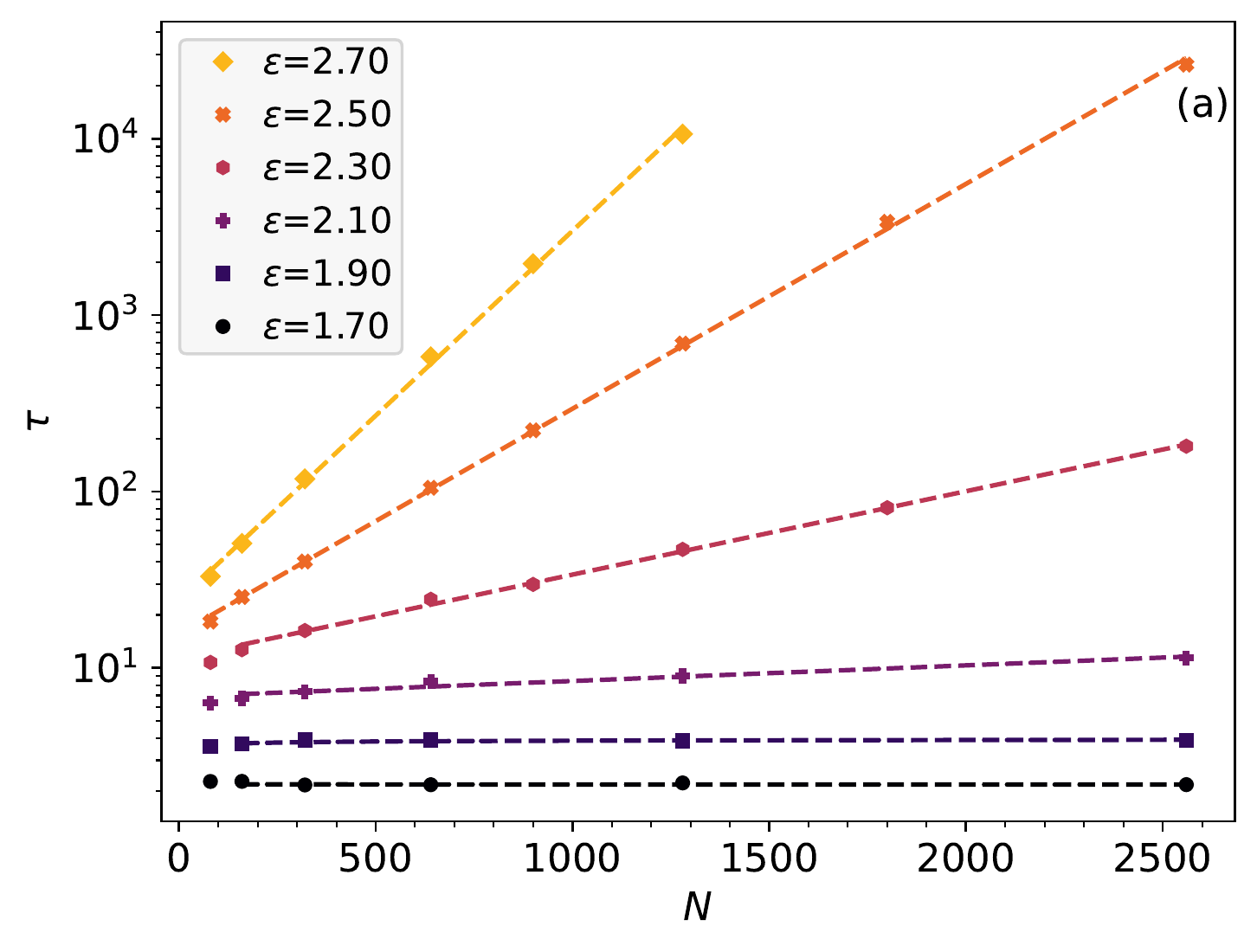}
    \label{fig:tau_vs_N}}
    \subfloat[]{\includegraphics[width=0.45\textwidth]{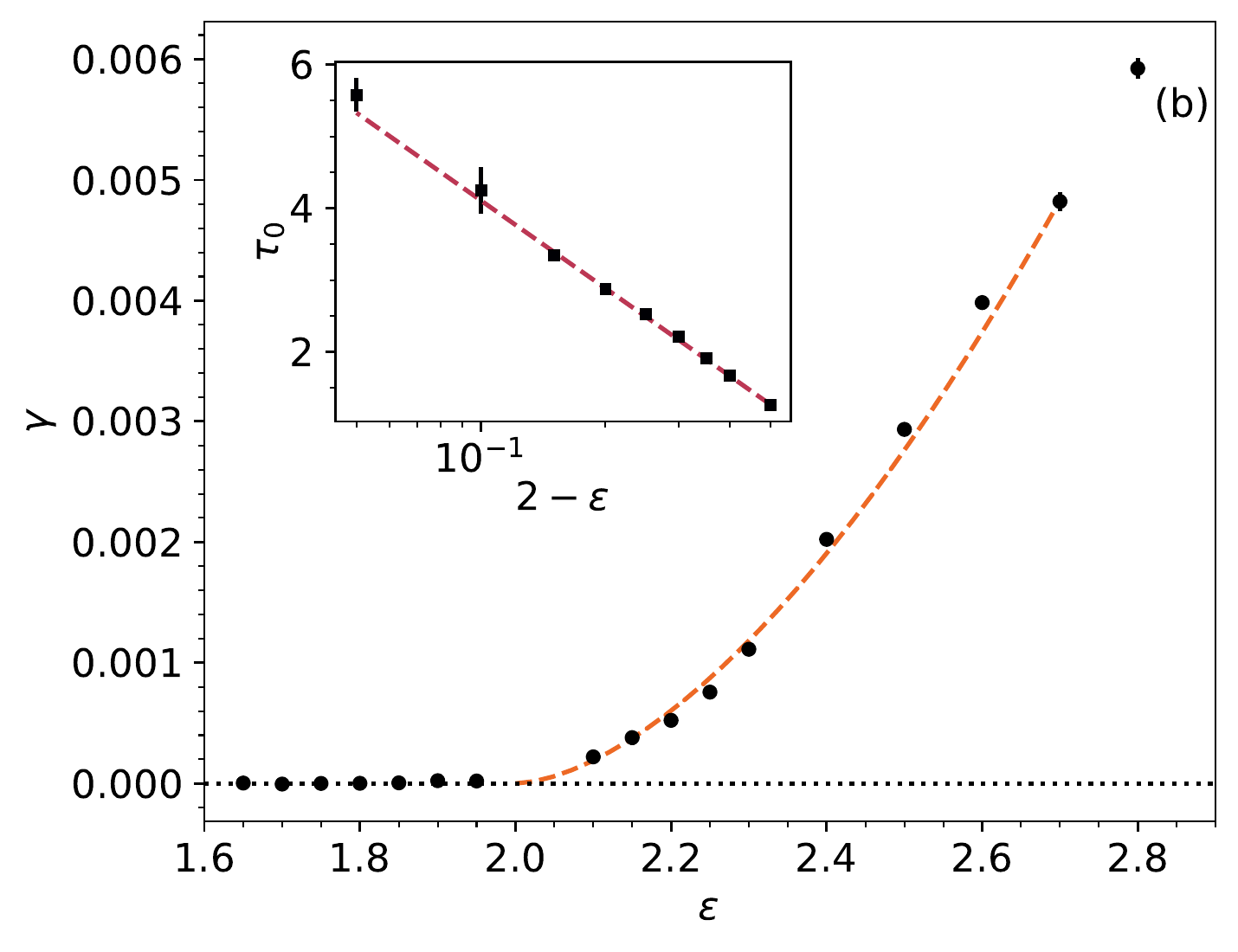}
    \label{fig:gamma_vs_eps}}
    \caption{(a) Correlation times $\tau$ extracted from the exponential decay of the correlation function, Eq.~\eqref{eq:Gq}: we can see that $\tau$ diverges in the thermodynamic limit as $\epsilon$ becomes greater than 2 (not all datasets are shown here to improve readability). We have also performed fits (dashed lines): for $\epsilon>2$, we employed $\log \tau = \gamma N + c$, from which we extracted the $\gamma$'s presented in the right panel. For $\epsilon<2$, instead, since $\tau$ is almost constant with $N$ we found that finite-size effects are well accounted for by the fitting function $\tau = w/ \log(N) + \tau_0$. The $\tau_0$'s obtained are displayed in the inset of the right panel. \\\hspace{\textwidth}
    (b) Exponent $\gamma$ as a function of $\epsilon$ (black dots). It can be clearly seen that $\gamma = 0$ within errors for $\epsilon < 2$, while $\gamma >0$ for $\epsilon > 2$. The orange, dashed line is a fit of the form $\log \gamma = \eta \log (\epsilon-2) + h$, yielding $\eta = 1.7 \pm 0.1$ and $h=-4.7 \pm 0.1$.
    \textit{Inset}: The value of $\tau \simeq \tau_0$ diverges logarithmically in the limit $\epsilon \to 2^-$ (see, for a comparison at the critical point, Fig.~\ref{fig:t1_vs_N}). The dashed line is a fit $\tau_0 = - \zeta \log(2-\epsilon) + u$, from which we find $\zeta = 1.76 \pm 0.02$ and $u = 0.05 \pm 0.01$. }
    \label{fig:fit_non_critical}
\end{figure}

In Fig.~\ref{fig:Gq} we show the time evolution of the (connected) correlation function
\begin{equation}
\label{eq:Gq}
    G(t) := \frac{1}{N} \sum_{i=1}^N \big\langle q_i(t+t') q_i(t') \big\rangle_{t'} - \frac{1}{N} \sum_{i=1}^N \big\langle  q_i(t+t') \big \rangle_{t'} \big\langle q_i(t') \big\rangle_{t'}
\end{equation}
where the angular brackets denote averaging wrt.\ the time variable in the subscript:
\begin{equation*}
    \big\langle A(t') \big\rangle_{t'} := \lim_{T_f \to \infty} \frac{1}{T_f} \int_0^{T_f} A(t') \, dt' .
\end{equation*}
At $t=0$, $G(0) = (\epsilon-1)$ because of the second constraint in Eq.~\eqref{eq:manifold}. Taking $t \to \infty$, instead, \emph{if the dynamics on the manifold is ergodic} it holds $G(\infty) = 0$. Conversely, if ergodicity is broken either because $\mathcal{M}_\epsilon$ is disconnected in pieces or because the dynamics is effectively confined in a smaller region, it holds $G(t)\to \mathit{const}>0$. According to the discussion before, $\mathcal{M}_\epsilon$ becomes disconnected for $\epsilon > N/2$. At this point there is a geometric obstruction to ergodicity: a trajectory starting in a neighborhood of, say, $q_1=O(N)$ cannot reach the neighborhood of any other $q_i=O(N)$ with $i \neq1$. Therefore, for these (large) values of $\epsilon$ the correlation function does not get to 0 as $t\to\infty$. Before then (viz.\ for any finite $N$, and any $\epsilon < N/2$) there is always a finite time scale $\tau$, after which the function $G(t)$ does get close to 0. This correlation time is a good proxy for an equilibrium time (since the charges $q_i$ are the only observables of the systems).

We also note that $G(t)$ must decay exponentially in $t$ (after, of course, a possible initial transient). This is due to the fact that a diffusion equation is associated to the Brownian motion:
\begin{equation}
\label{eq:FP}
    \partial_t P(q,t) = \frac{1}{2} \Delta P(q,t),
\end{equation}
where the Laplacian has the usual definition in curvilinear coordinates $\Delta:=g^{-1/2}\partial_a (g^{1/2} g^{ab}\partial_b)$. The smallest eigenvalue of $-\Delta$ is $\lambda_0=0$, and the corresponding (properly normalized) eigenvector is nothing but the uniform (microcanonical) distribution $P(q,t\to\infty)=\phi_0(q)$. Since the Laplacian on a compact Riemannian manifold has a pure point spectrum, the first eigenvalue $\lambda_1>0$ and the gap, which we will denote as $\lambda_1=:1/\tau,$ controls the asymptotic decay of $P(q,t)\simeq \phi_0(q)+c_1 \phi_1(q)e^{-t/\tau}+...$. In particular, this means that $G(t)\sim e^{-t/\tau}$ for large $t$, as claimed before.

The exponential decay can be seen clearly in Fig.~\ref{fig:Gq_nonCritical}. Within the exponential form, one can distinguish two cases: for $\epsilon < 2$ the curves fall approximately on a universal curve, which is the limit $N \to \infty$; for $\epsilon \geq 2$ various system sizes have different decays: larger systems decay on a \emph{longer} timescale and there is no obvious limit $N \to \infty$. This corresponds to the following statement on the spectrum of the Laplacian: for $\epsilon<2$ the gap remains finite when $N\to\infty$, while for $\epsilon\geq 2$ the gap closes with $N$. We find numerically that the gap closes exponentially with $N$: $\tau \simeq e^{\gamma N}$ for some rate $\gamma>0$, see Fig.~\ref{fig:fit_non_critical}. This fact implies that $\epsilon=2$ is a \emph{dynamical} critical point, at which the dynamics becomes scale-invariant \cite{tauber2014}.

More precisely, for $\epsilon<2$ the timescale $\tau$ is constant with $N$ and grows with $\epsilon$, ultimately diverging logarithmically as $\epsilon\to 2^-$ (see the inset of Fig. \ref{fig:gamma_vs_eps}). For $\epsilon>2$, it is $\gamma(\epsilon)$ which grows with $\epsilon$: $\gamma \sim (\epsilon-2)^\eta$ with $\eta \simeq 1.7\pm 0.1$ (see Fig.~\ref{fig:gamma_vs_eps}).

\begin{figure}[t]
    \centering
    \subfloat[]{\includegraphics[width=0.45\textwidth]{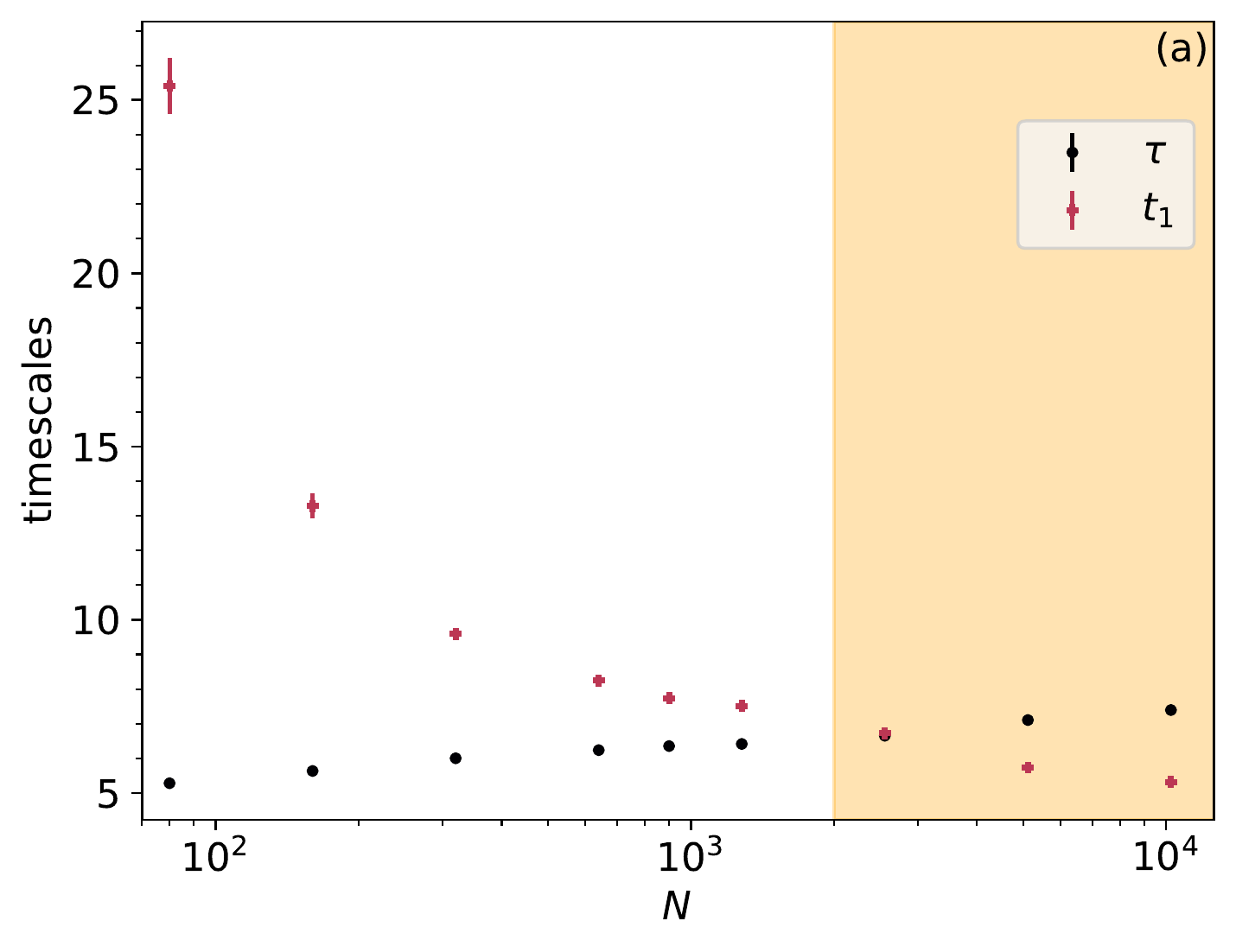}
    \label{fig:t1_vs_N}}
    \subfloat[]{\includegraphics[width=0.45\textwidth]{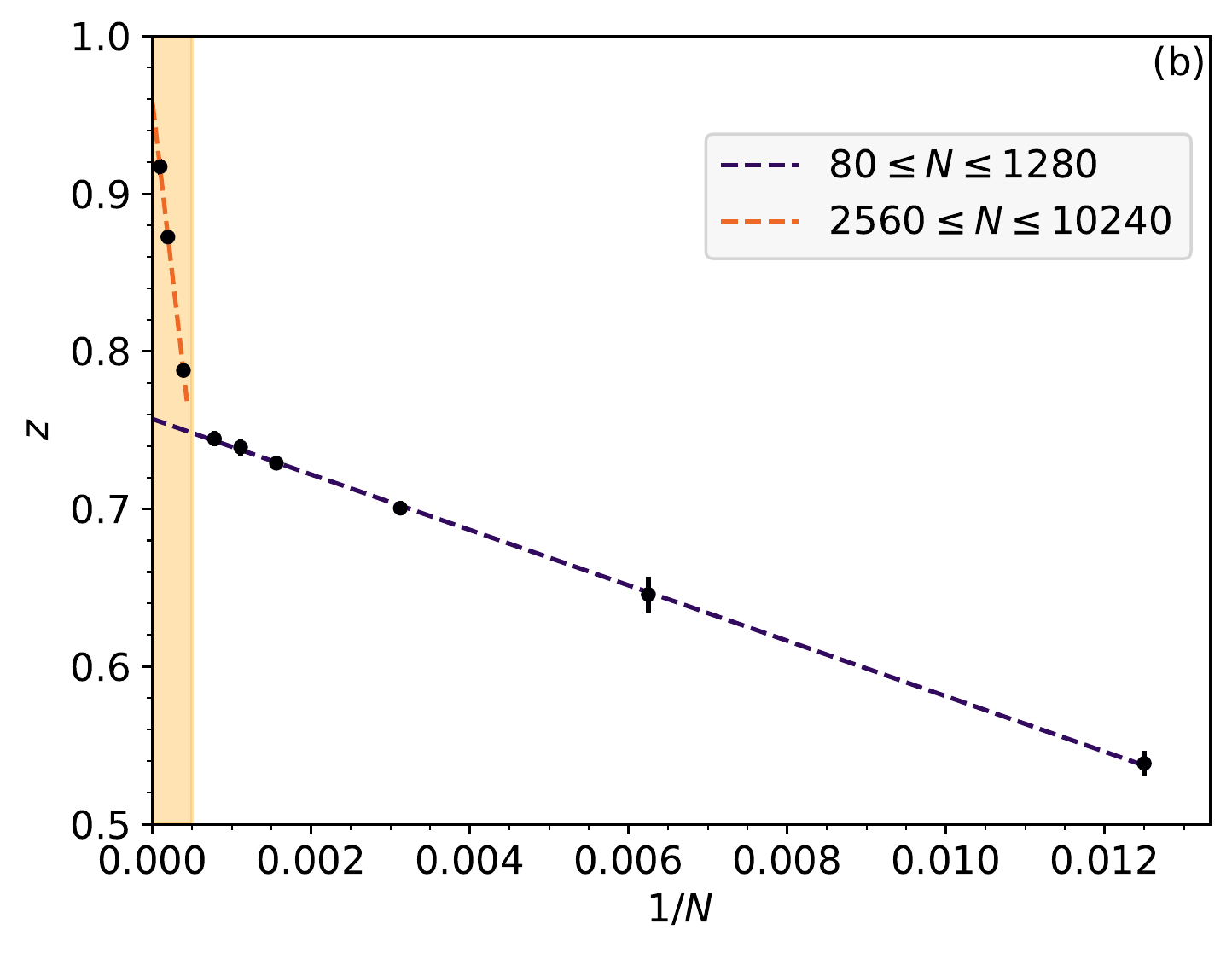}
    \label{fig:z_vs_N}}
    \caption{(a) Typical timescales of the exponential ($\tau$) and power-law ($t_1$) decay of the correlation function at the critical point $\epsilon=2$, found from the fits of the correlation function $G(t)$ (Fig.~\ref{fig:Gq_critical}). We see that the crossover to power-law decay takes place at $N \gtrsim 2000$ (shaded region), so much larger system sizes are needed to extract a clean dynamical critical exponent $z$. \\\hspace{\textwidth}
    (b) Corresponding values of the dynamical critical exponent $z$. The errorbars represent the fit errors, that surely underestimate the strong fluctuations at finite $N$ (see also \cite{Gradenigo2021Localization,Gradenigo2021Condensation}). Therefore, we can only present two possible fits, one excluding the smallest $N$ points and one the largest, and give a value of $z$ respectively $z\simeq 1.0$ and $z\simeq 0.75$. Since the smallest $N$ points have $t_1>\tau$ (as seen in panel (a)), which is a clear pre-asymptotic behavior, we would tend to discard them in favour of the 3 largest $N$ points in the dataset.}
    \label{fig:fit_critical}
\end{figure}

To sum up: to the left of the critical point $\epsilon=2$ the dynamics is ergodic and $G(t)\to 0$ for any $N$ \emph{and also for the limit} $N\to\infty$; to the right, instead, the relaxation becomes progressively slower as $N$ increases and in the limit $N\to\infty$ it holds $G(t)\to \mathit{const} > 0$. The limiting functional form at $\epsilon=2$ must be a function decaying to 0, but slower than an exponential. We checked that, in a small right neighborhood of $\epsilon=2$, the fitting function $G(t) = (\epsilon-1) e^{-t/\tau} / (1 + (t/t_1)^z)$ works pretty accurately with $z\in [0.7,1.0]$ depending on the values of $N$ used (see Figs.\ \ref{fig:Gq_critical} and \ref{fig:fit_critical} for the details). Indeed, for this estimate to be useful one must have $t_1\ll \tau\to\infty,$ to ensure a sufficiently large fitting window. Since $\tau$ grows with $N$ (albeit only logarithmically) while $t_1$ decreases (see Fig.~\ref{fig:t1_vs_N}), it will eventually hold $t_1 \ll \tau$. Unfortunately, this crossover takes place roughly at the largest system sizes we were able to simulate, so the values of $z$ we can extract cannot be considered precise. By using only the points in the ``asymptotic region" (shaded region in Fig.~\ref{fig:fit_critical}), $z\simeq 1.0$, while smaller values of $N$ (non-shaded region) we have a considerably smaller $z\simeq 0.75$. The precise values of the critical exponents clearly requires further numerical investigations. We end by noticing that the form of $G(t)$ at criticality can be related to the distribution $\rho(\lambda)$ of the eigenvalues of the Laplacian at $\epsilon=2$, which must be of the form $\rho(\lambda) \sim \lambda^{z-1}$ near $\lambda = 0$.

%------------------------------------------CONCLUSIONS------------------------------------------%  

\section{Conclusions and Outlook}
\label{sec:conclusions}

We have studied the mechanism for weak ergodicity breaking at high energy densities in a fully connected DNLSE model. We have shown that, whatever the interactions between sites (kinetic energy term) are, they can be neglected for $\epsilon \geq \epsilon_s=1.481...$ that corresponds to a finite temperature $T_s = 2g = 2$. We are left therefore with a purely potential model, whose physical properties reflect the geometrical properties of the potential energy surface and therefore are subject to a localization transition at infinite temperature (corresponding to $\epsilon_c=v=2)$. After proving that the microcanonical, potential energy surface is connected for all extensive energies (therefore energy densities of $O(1)$), we show that the localization transition is due to a phase transition in the order parameter $\gamma=-(\log \lambda_1)/N$, where $\lambda_1$ is the smallest non-zero eigenvalue of $-\Delta$, the Laplacian on the (curved) equipotential surface. For $\epsilon<2$ we have $\gamma=0$ and for $\epsilon>2$ we have $\gamma\sim (\epsilon-2)^\eta$ with $\eta$ around 2. This puts on firmer ground the connection between the works on thermodynamics (like \cite{Gradenigo2021Localization,Gradenigo2021Condensation}) and those on the dynamics (like \cite{Iubini2013Discrete,Iubini2014Coarsening,gotti2020finitesize}). The approximation in which one can neglect the kinetic energy is exact on the fully connected model, and one can imagine that it is a good approximation for a finite-dimensional lattice, therefore making our results qualitatively compelling for the experimental observations in \cite{eiermann2004bright,bloch2008many} and the numerical works \cite{Flach2018Weakly}. Making a quantitative connection, and computing a possible $1/\kappa$ series of corrections to our results is left for future work.

The transition taking place at $\epsilon_c=2$ makes the equilibration time $\tau$ change from $O(1)$ to exponentially large in $N$, $\tau\sim e^{\gamma N}$: the phenomenology is very similar to that observed in the MBL-like phase of Josephson junction arrays \cite{pino2016nonergodic,pino2017multifractal}, and of quantum glasses as well \cite{laumann2014many,baldwin2016many,mossi2017ergodic,baldwin2017clustering}. However, the nature of the transition in the DNLSE seems to be of entropic origin: the volume of the region of phase space around any given localized configuration is exponentially larger than the volume connecting two localized configurations, therefore making the passage from one localized configuration to another exponentially unlikely. Quantum mechanical localization, in contrast, is a consequence of interference and it vanishes when $\hbar\to 0$. It is also tempting to notice that the lowest eigenvalue of $-\Delta$ becoming exponentially small in a large parameter is precisely what happens in localized quantum-mechanical Schr\"odinger equations. However, given these elements, we cannot argue more than a similarity at a formal level. 

We also refrain from speculating on the effect of turning on $\hbar$. Previous works have shown that, for $\hbar \neq 0$ and at least in a 1d geometry, transport is strongly suppressed as $T \to \infty$ \cite{DeRoeck2014Asymptotic} and non-Gibbs state exist for $\epsilon \geq 2$ as well \cite{Cherny2019NonGibbs}. In a more general setting, on the one hand one would expect that a charge localized on a site could tunnel quantum-mechanically towards a neighboring site; on the other hand the effects of interference should be taken into account as in \cite{pino2016nonergodic,pino2017multifractal}. Sorting out the leading effects of quantization upon the system \eqref{eq:Hamiltonian}, at least at the semiclassical level, is left out for future work.

%------------------------------------------ACKNOWLEDGEMENTS------------------------------------------%  
\begin{acknowledgements}
We would like to thank Sergio Caracciolo, Rosario Fazio, Sergej Flach, Giacomo Gradenigo, Giorgio Parisi, and Federico Ricci-Tersenghi for discussion. C.V.\ thanks also ICTP for hospitality during the initial part of this work. This work is supported by the Trieste Institute of Quantum Technolgies (TQT).
\end{acknowledgements}

%------------------------------------------BIBLIOGRAPHY------------------------------------------%  
\bibliographystyle{spphys} 
\bibliography{biblio}

\appendix

%------------------------------------------APPENDIX-TOPOLOGY------------------------------------------%  
\section{The topology of the equipotential manifolds}
\label{app:topology}

Here, we want to provide a detailed description of the topology of the manifolds $\mathcal{M}_\epsilon$ determined by equation \eqref{eq:manifold}, as outlined in Sec.\ \ref{sec:topology} in terms of Morse theory. Actually, in order to avoid the technicalities of stratified Morse theory, we prefer to adopt a different, more direct approach, using only basic notions and results of piecewise topology, for which we refer to \cite{RourkeSanderson}.

According to \eqref{eq:manifold}, we can think of $\mathcal{M}_\epsilon$ as the intersection between the sphere $\mathbb{S}^{N-1}_\epsilon \subset \mathbb{R}^N$ centered at the origin of radius $\sqrt{\epsilon N}$ and the simplex $\Delta^{N-1} \subset \mathbb{R}^N$ affinely spanned by the vectors $Ne_1, \dots, Ne_N$, where $e_1,\dots, e_N$ is the canonical base of $\mathbb{R}^N$. 

Given any $k$-dimensional sub-simplex $\Sigma = \langle N e_{i_1}, \dots, N e_{i_{k+1}}\rangle \subset \Delta^{N-1}$ with $0 \leq k \leq N-1$, let $b(\Sigma) = N
(e_{i_1} + \dots + e_{i_{k+1}})/(k+1)$ be the barycenter of $\Sigma$. We indicate by $\Gamma$ the barycentric subdivision of $\Delta^{N-1}$, whose $k$-simplices are given by $\langle b(\Sigma_{i_1}), \dots, b(\Sigma_{i_{k+1}})\rangle$ for any ascending chain $\Sigma_{i_1} \subset \dots \subset \Sigma_{i_{k+1}}$ of sub-simplices of $\Delta^{N-1}$. Moreover, for every $0 \leq k \leq N-1$, let $\Gamma_k$ and $\Gamma^k$ denote the sub-complexes of $\Gamma$ consisting of all simplices $\langle b(\Sigma_{i_1}), \dots, b(\Sigma_{i_{\ell+1}}) \rangle$ such that $\dim \Sigma_{i_j}\leq k$ and $\dim \Sigma_{i_j} \geq k+1$, respectively.

We observe that $\Gamma_k$ coincides with the barycentric subdivision of the $k$-skeleton $\Delta^{N-1}_k$ of $\Delta^{N-1}$, hence $\dim{\Gamma_k}=k$, while $\Gamma^k$ is the sub-complex of $\Gamma$ consisting of all the simplices that are disjoint from $\Delta^{N-1}_k$, and $\dim{\Gamma^k} = N - k - 2$. Furthermore, $\Gamma$ can be expressed as the affine join $\Gamma = \Gamma_k \ast \Gamma^k$ in $\mathbb{R}^N$, that is for every simplex $\Sigma \in \Gamma$ we have $\Sigma = (\Sigma \cap \Gamma_k) \ast (\Sigma \cap \Gamma^k)$. Then, there is a well-defined pseudo-radial projection $\pi_k:\Gamma - \Gamma^k \to \Gamma_k$, which collapses $\Sigma - \Gamma^k$ to $\Sigma \cap \Gamma_k$ for every simplex $\Sigma$ of $\Gamma$.

Now, we can start our description of $\mathcal{M}_\epsilon$. First of all, we note that $\mathcal{M}_\epsilon$ is empty for $\epsilon < 1$ and $\epsilon > N$, while it consists of the single point $b(\Delta^{N-1})$ for $\epsilon = 1$ and of the $N$ vertices of $\Delta^{N-1}$ for $\epsilon = N$. 

When $\epsilon$ ranges in the interval $[1,N]$ the sphere $\mathbb{S}^{N-1}_\epsilon$ meets transversally each subsimplex of $\Delta^{N-1}$, except for $\epsilon = N/k$ with $k=1,\dots,N$, in which case $\mathbb{S}^{N-1}_\epsilon$ is tangent to all the $\binom{N}{k}$ subsimplices of $\Delta^{N-1}$ of dimension $k-1$ at their barycenters.
As a consequence, each $\mathcal{M}_\epsilon$ can be endowed with a structure of stratified space, whose strata are the components of the intersections of $\mathbb{S}^{N-1}_\epsilon$ with the open simplices of $\Delta^{N-1}$. Moreover, such structure is the same up to smooth isomorphism for all $\epsilon$ in each open interval $(N/(k+1),N/k)$ with $k=1,\dots,N-1$.

Now, fix $\epsilon \in (N/(k+1),N/k)$ with $k=1,\dots,N-1$. In order to describe $\mathcal{M}_\epsilon$, we consider the affine subspace $A^{N-1} \subset \mathbb{R}^N$ spanned by $N e_1, \dots, N e_N$, the $(N-2)$-dimensional sphere $\mathbb{S}^{N-2}_\epsilon = \mathbb{S}^{N-1}_\epsilon \cap A^{N-1}$, and the $(N-1)$-dimensional closed ball $B^{N-1}_\epsilon = B^N_\epsilon \cap A^{N-1}$ bounded by $\mathbb{S}^{N-2}_\epsilon$ in $A^{N-1}$, where $B^N_\epsilon \subset \mathbb{R}^N$ is the $N$-cell centered at the origin of radius $\sqrt{\epsilon N}$.

The inequality $\epsilon > N/(k+1)$ implies that $\Gamma^{k-1} \subset \mathrm{Int}\, B^{N-1}_\epsilon$, since all the vertices of $\Gamma^{k-1}$ belong to $\mathrm{Int}\, B^N_\epsilon$. On the other hand, the inequality $\epsilon < N/k$ implies that $\Gamma_{k-1} \subset A^{N-1} - B^{N-1}_\epsilon$, being $d(0,\Sigma)=\|b(\Sigma)\| = N/k$ for any $(k-1)$-dimensional face $\Sigma$ of $\Delta^{N-1}$. Therefore, $\mathbb{S}^{N-2}_\epsilon$ transversally meets in a single point each segment $\langle p,q\rangle \subset \Sigma$ with $p \in \Gamma_{k-1}$, $q \in \Gamma^{k-1}$ and $\Sigma$ a simplex of $\Gamma$. Hence, $\mathcal{M}_\epsilon$ is pseudo-radially equivalent, to the boundary $\mathrm{Bd}\, N(\Gamma_{k-1},\Gamma)$ of a regular neighborhood $N(\Gamma_{k-1},\Gamma)$ of $\Gamma_{k-1}$ in $\Gamma$. Finally, due to the inclusion $\Gamma_{k-1} \subset \mathrm{Bd}\,\Gamma$, we can conclude that $\mathcal{M}_\epsilon$ is topologically equivalent to a regular neighborhood $N(\Gamma_{k-1}, \mathrm{Bd}\,\Gamma)$ of $\Gamma_{k-1}$ in $\mathrm{Bd}\,\Gamma$. Notice that $N(\Gamma_{k-1}, \mathrm{Bd}\,\Gamma)$ coincides with the suplevel set $M_\epsilon(\varphi)$ considered in Sec.\ \ref{sec:topology}.

Of course, up to radial projection in $A^{N-1}$ centered at $b(\Delta^{N-1})$, we can identify $\mathrm{Bd}\,\Delta^{N-1}$ with $\mathbb{S}^{N-2}_\epsilon$ and $\Gamma_{k-1}$ with a sub-complex $\Gamma_{k-1,\epsilon} \subset \mathbb{S}^{N-2}_\epsilon$, in such a way that $\mathcal{M}_\epsilon$ turns out to be topologically equivalent to a regular neighborhood $N(\Gamma_{k-1,\epsilon},\mathbb{S}^{N-2}_\epsilon)$ of $\Gamma_{k-1,\epsilon}$ in $\mathbb{S}^{N-2}_\epsilon$. Then, there is a collapse $\mathcal{M}_\epsilon \searrow \Gamma_{k-1,\epsilon} \cong \Gamma_{k-1}$.

The argument above also applies to the case of $\epsilon = N/k$, with the only difference that in this\break case the regular neighborhoods $N(\Gamma_{k-1}, \mathrm{Bd}\,\Gamma)$ and $N(\Gamma_{k-1,\epsilon},\mathbb{S}^{N-2}_\epsilon)$ are relative to the $0$-dimen\-sional subcomplex $\{b(\Sigma)\,|\, \Sigma \text{ is a ($k-1$)-face of $\Delta^{N-1}$}\}$, but still we have $\mathcal{M}_\epsilon \searrow \Gamma_{k-1,\epsilon} \cong \Gamma_{k-1}$.

Summarizing, $\mathcal{M}_1$ consists of a single point and 
$\mathcal{M}_\epsilon$ is homotopically equivalent to $\Delta^{N-1}_{k-1}$ for every $N/(k+1) < \epsilon \leq N/k$ and $k=1,\dots,N-1$. In particular, $\mathcal{M}_\epsilon$ is connected non-empty if $1 \leq \epsilon \leq N/2$, while it splits into the disjoint union of $N$ disks, one for each vertex of $\Delta^{N-1}$ if $N/2 < \epsilon \leq N$.

%------------------------------------------APPENDIX-NUMERICS------------------------------------------%  

\section{Numerical implementation of the Brownian motion}
\label{app:numerics}

To simulate efficiently a Brownian dynamics on the manifold $\mathcal{M}_\epsilon$ of Eq.~\eqref{eq:manifold} we proceeded as follows. As a first thing, we changed variables to
\begin{equation*}
    x_i := q_i - 1,
\end{equation*}
in view of the fact that $\langle q_i(t) \rangle_t=1$ if the dynamics is ergodic. In term of the $x$ variables, Eq.~\eqref{eq:manifold} reads
\begin{numcases}{}
    \label{eq:constraint_x_lin} \textstyle
    \sum_{i=1}^N x_i = 0 \\
    \label{eq:constraint_x_quad} \textstyle
    \sum_{i=1}^N x_i^2 = N\epsilon \\
    \label{eq:constraint_x_pos} \textstyle
    x_i \geq -1 \quad \forall i=1,2,\dots,N.
\end{numcases}

Then, at each time step $t$ we draw $N$ iid. gaussian variables $dW_i(t) \sim \mathcal{N}(0,dt)$ and propose a move $x_i(t) \mapsto y_i(t) := x_i(t) + dW_i(t)$. The point $\vec{y}$ lies no more on $\mathcal{M}_\epsilon$, since we have violated the constraints in Eqs.~\eqref{eq:constraint_x_lin}--\eqref{eq:constraint_x_pos} with probability 1. Therefore, we need the following passages:
\begin{enumerate}
    \item we impose Eq.~\eqref{eq:constraint_x_lin} by simply subtracting $\frac{1}{N}\sum_j y_j$ from each $y_i$, obtaining a set of $y_i'$;
    \item we impose Eq.~\eqref{eq:constraint_x_quad} by multiplying each $y_i'$ by $N\epsilon / \sum_j (y_j')^2$, obtaining a set of $y_i''$.
\end{enumerate}
Now both Eqs.~\eqref{eq:constraint_x_lin} and \eqref{eq:constraint_x_quad} are satisfied, but it may be that some inequality in Eq.~\eqref{eq:constraint_x_pos} is violated. Therefore, we need the last passage:
\begin{enumerate}
\setcounter{enumi}{2}
    \item we check that each $y''_i \geq -1$, and if it is not the case we reflect $y''_i \mapsto y'''_i = -2 - y_i''$.
\end{enumerate}
Now all the constraints in Eq.~\eqref{eq:constraint_x_pos} are satisfied, but we are violating again Eqs.~\eqref{eq:constraint_x_lin} and \eqref{eq:constraint_x_quad}. Thus, we start again from point (i) and repeat the procedure until every constraint is satisfied, finally obtaining a point $\vec{x}(t+dt)$. Typically, a few iterations are sufficient. We have also explicitly checked that we do not introduce jumps, and that the statistics of $\| \vec{x}(t+dt) - \vec{x}(t) \|$ is very close to that of $\| d\vec{W} \|$.

\end{document}